\documentclass[aps,prb,reprint,superscriptaddress,nofootinbib]{revtex4-2}
\usepackage{graphicx}
\usepackage{amssymb}
\usepackage{amsmath}
\usepackage{color}
\usepackage{enumerate}
\usepackage{bm}
\begin{document}

\title{Nonlinear and Chaotic Dynamics of Two Dipolar-Coupled Anisotropic Spins under Periodic Excitation}
%\title{Kinetic Regimes and Scaling Laws in Pulse-driven Accumulative Switching of HZO}
\author{Manish Anand}
\email{itsanand121@gmail.com}
%\altaffiliation{These authors contributed equally to this work}
\affiliation{Department of Physics, Bihar National College, Patna University, Patna-800004, India.}

\date{\today}

\begin{abstract}
We investigate the non-linear magnetization dynamics of two identical anisotropic macrospins coupled by dipolar interaction and subjected to periodic transverse excitation. The coupled Landau-Lifshitz-Gilbert equations are solved numerically using a fourth-order Runge-Kutta method. By combining bifurcation diagrams, local-maxima analysis, Lyapunov exponents, and two-dimensional stability maps, we characterize how dipolar coupling, transverse driving amplitudes, and uniaxial anisotropy govern the long-time dynamics. The results reveal a non-monotonic reorganization of dynamical regimes, with alternating intervals of enhanced and suppressed instability. Analysis of symmetric and antisymmetric coordinates distinguishes the collective and relative spin responses, while Fourier spectra and three-dimensional trajectories provide complementary frequency-domain and geometrical characterizations. Simultaneous positivity of the two largest Lyapunov exponents identifies parameter regions consistent with hyperchaotic dynamics. Finally, analytical linearization and ferromagnetic-resonance calculations reveal the distinct coupling dependence of the collective-mode frequencies and clarify the limitations of the linearized description beyond the stability thresholds of the collinear reference state. These findings establish a unified framework for understanding the interplay of dipolar coupling, anisotropy, and periodic excitation in the non-linear dynamics of a minimal interacting-spin system.
\end{abstract}

\maketitle
\newpage
\section{Introduction}

%The understanding of complex magnetization dynamics of magnetic systems is not only crucial from theoretical point of views but also due to its  pivotal role in modern spintronics, magnonics, and microwave technologies~\cite{bibes2026,rana2024,kulbak2026,serha2026,koopmans2005,tserkovnyak2005,barman2020,choe2004vortex}. The understanding of such dynamics of interacting spins could provide a  fundamental framework for understanding the interplay between magnetic interactions, anisotropy, dissipation, and external excitation, which is crucial for designing spintronics and magnonics-based devices. 

The understanding of complex magnetization dynamics of magnetic systems is important not only from a fundamental physics perspective but also because of its pivotal role in modern spintronics, magnonics, and microwave technologies~\cite{bibes2026,rana2024,kulbak2026,serha2026,koopmans2005,tserkovnyak2005,barman2020,choe2004vortex}. In particular, investigating interacting spins provides fundamental insights into how magnetic coupling, anisotropy, dissipation, and external excitation collectively govern magnetization dynamics. Such insights are essential for understanding collective magnetic behaviour and may guide the design and optimization of spintronics and magnonic devices. The magnetization dynamics of an isolated spin are governed by the effective magnetic field, magnetic anisotropy, and damping, whereas interactions between spins introduce additional coupling mechanisms in a system of interacting spins that enrich the dynamical behaviour~\cite{tatsumi2025chaotic,laroze2011characterization}. Since the effective field acting on each spin in such interacting system depends on the instantaneous orientations of the other spins, their coupled evolution can exhibit complex non-linear behaviour. Under periodic excitation, these non-linearities can give rise to a variety of dynamical responses, ranging from regular periodic motion to complex oscillations and deterministic chaos~\cite{barrio2012topological,elsayed2014signatures,ermakov2025classical,lebel2022chaos}. Understanding the physical mechanisms governing these transitions is therefore essential for elucidating the non-linear dynamics of interacting magnetic systems.

The Landau-Lifshitz-Gilbert (LLG) equation provides the standard description of magnetization dynamics by incorporating precessional motion around the effective field and dissipative relaxation~\cite{botha2026landau,xu2023unified}. For interacting magnetic moments subjected to time-dependent fields, the effective field varies with both the instantaneous magnetization configuration and the external excitation. Consequently, even when the applied external excitation field contains a single frequency, the magnetization response can develop additional frequency components and exhibit non-trivial changes in its long-time behaviour~\cite{taniguchi2019chaos,rezende1995chaotic}. The resulting dynamics can not be characterized adequately from the amplitude of the magnetization or its frequency spectrum alone. A combined analysis of the trajectories, their stability, and their frequency content is required to distinguish regular responses from chaotic dynamics and to understand how the transitions between these regimes depend on the system parameters.

Among the interactions governing coupled magnetic moments, dipolar interaction is distinguished by its dependence on the relative orientations of the moments and their spatial separation~\cite{laroze2008dynamics,cortes2008dynamical}. Its directional character makes the interaction intrinsically anisotropic: the effective field generated by one magnetic moment acts differently along different spatial directions~\cite{anand2016spin}. When dipolar interaction is combined with uniaxial magnetic anisotropy and periodic transverse excitation, it modifies the torque experienced by each moment and changes the balance between precession, damping, and driving~\cite{collings2023generalized,kalmykov2020dipole}. Varying the interaction strength can therefore influence not only the characteristic frequencies of the system but also the stability and structure of its non-linear trajectories. The resulting response need not evolve monotonically with increasing coupling, since changes in the effective-field configuration can either enhance or suppress dynamical instability over different parameter intervals.

In these contexts, there are various works worth mentioning. Ferona and Camley investigated a uniformly magnetized ellipsoidal particle under static and oscillatory linearly and circularly polarized excitation using numerical solutions~\cite{ferona2017nonlinear}. They observed complex magnetization switching, prolonged transients, quasi periodicity, and chaos. Tomohiro Taniguchi {\it et al.} investigated synchronized, periodic, and chaotic dynamics in spin torque oscillator with two free layers~\cite{taniguchi2019}. They observed different dynamical regimes, including synchronization, auto-oscillation at different frequencies, and chaotic dynamics, depending on the applied current and magnetic field. Urzagasti {\it et al.} studied spin dynamics of two interacting spins using numerical simulations~\cite{urzagasti2015}. They reported multiple transitions between regular and chaotic behaviour depending on the control parameters. V\'elez {\it et al.} analyzed the magnetization dynamics of an anisotropic magnetic particle~\cite{velez2020}. They observed islands of synchronization for the magnetization and the driven field and several shrimp structures with different periods. In a similar study, Bragard {\it et al.} reported multiple transitions between regular and chaotic behaviours as a function of external excitation~\cite{bragard2011}. M.~Lakshmanan investigated the dynamics of single spin using analytical techniques and numerical simulations in a greater detail~\cite{lakshmanan2011}. They reported fascinating magnetic structure such as bifurcation, periodic and chaotic states as a function of external excitation. The work of Serpico {\it et al.} on nonlinear magnetization dynamics in nanomagnets reviews the equilibrium configurations, phase portraits, precessional motion, damping, switching, and resonance of uniformly magnetized particles~\cite{Serpico2007}. 

Despite extensive studies of non-linear magnetization dynamics in driven single-spin system and coupled magnetic structures, how anisotropic dipolar coupling reshapes the transition between regular and chaotic dynamics in a minimal two-spin system remains an important question. In particular, the interplay between dipolar coupling, magnetic anisotropy, and transverse periodic excitation requires a unified analysis that connects the stability of collective modes with non-linear bifurcations and the emergence of chaos. The distinction between symmetric and antisymmetric responses can be especially useful for revealing how coupling influences the collective and relative motion of the spins beyond the linear-response regime. In contrast to larger spin assemblies, a two-spin system permits the dynamics of each moment to be examined directly while retaining the essential feedback associated with magnetic coupling. In the present work, we consider two identical anisotropic spins positioned along the \(x\)-axis, with their easy axes aligned along the \(z\)-direction. A static magnetic field is applied along the $z$-axis, while a transverse periodic field drives the system. In such a geometry, the dipolar interaction contributes differently to the \(x\)-, \(y\)-, and \(z\)-components of the effective field. In particular, the transverse components are not equivalent from the perspective of the interaction. As a result, varying the amplitudes of the \(x\)- and \(y\)-directed driving fields can produce distinct dynamical responses, even though both components have the same sinusoidal time dependence. This geometry therefore allows the coupled effects of dipolar interaction, transverse driving, and uniaxial anisotropy to be investigated within a single model.

An important aspect of the two-spin system is the distinction between collective motion and relative motion, which can be examined by introducing symmetric and antisymmetric coordinates constructed from the sum and difference of the two magnetization vectors, respectively. The symmetric coordinates describe the combined, predominantly in-phase response, whereas the antisymmetric coordinates characterize differences between the spin trajectories and are associated with predominantly out-of-phase motion~\cite{lee2012current}. Such a representation provides a useful means of determining whether changes in the magnetization dynamics are expressed primarily in the collective response or in the relative motion of the two spins~\cite{patchett2022symmetry}. It also facilitates the examination of the growth of perturbations initially oriented along symmetric and antisymmetric directions. On the other hand, the analytical linearization of the governing LLG equation around the collinear configuration provides a reference for understanding the collective excitations in the small-amplitude limit. This analysis yields distinct symmetric and antisymmetric mode frequencies, with different dependences on the dipolar coupling strength. The resulting analytical expressions can clarify how the interaction modifies the characteristic precessional timescales of the two collective modes and provide a basis for comparison with the numerical ferromagnetic-resonance (FMR) response~\cite{edathumkandy2025}. 
%However, the linearized description is restricted to the assumed reference configuration and cannot, by itself, characterize the stability of finite-amplitude, periodically driven trajectories. A combined analytical and numerical investigation is therefore necessary to connect the collective-mode structure with the non-linear dynamical response.

In this work, we investigate the non-linear dynamics of two dipolar-coupled anisotropic macrospins governed by the Landau–Lifshitz–Gilbert equation under periodic transverse excitation. By combining bifurcation diagrams, Poincaré sections, Lyapunov exponents, Fourier spectra, and ferromagnetic-resonance analysis, we systematically examine how dipolar coupling, excitation field amplitudes, and anisotropy shape the dynamical response. Our central objective is to establish how the dipolar coupling strength, the two transverse driving amplitudes, and the uniaxial anisotropy jointly control the organization and stability of the long-time magnetization dynamics. Therefore, treating these parameters independently, we examine their combined influence on the effective fields and on the resulting spin trajectories. Two-dimensional Lyapunov maps are further used to examine the dependence of dynamical stability on combinations of control parameters. In addition, frequency-dependent FMR calculations are compared with the analytical collective-mode frequencies to connect the linear response with the interaction-dependent dynamical behaviour.

The different diagnostics employed in the present work provide complementary information about the system. Poincaré sampling and local-maxima analysis reveal changes in the organization of the long-time response as parameters are varied, while Lyapunov exponents quantify sensitivity to small perturbations and provide a stability-based criterion for identifying deterministic chaos~\cite{sprott2003chaos}. The Fourier spectra characterize the distribution of spectral weight among the frequencies generated by the non-linear motion, and the three-dimensional trajectories provide a geometrical representation of the evolution of the individual spins and collective coordinates~\cite{baker2017proposal,saavedra2021effect}. The FMR calculations address the frequency-dependent response and complement the non-linear parameter scans performed at a fixed driving frequency. In particular, the combined use of these methods allows complex magnetization responses to be assessed using both geometrical and quantitative stability information, rather than inferring chaos from spectral broadening or trajectory complexity alone.

The present study attempts to establish a unified description of how anisotropic dipolar coupling, uniaxial anisotropy, and periodic excitation govern the non-linear response of a minimal interacting-spin system. By comparing individual-spin dynamics with symmetric and antisymmetric coordinates, we identify the distinct contributions of collective and relative motion to the parameter-dependent response. The analytical normal-mode calculation provides a reference for interpreting the interaction-dependent FMR features, while the non-linear dynamical diagnostics characterize the stability and complexity of the driven trajectories beyond the linear regime. Together, these analyses clarify why the emergence of dynamical instability depends on the combined values of the interaction, driving amplitudes, and anisotropy, rather than on a single control parameter.

The remainder of the paper is organized as follows. The model and mathematical formulation are presented first, including the dimensionless LLG equation, the collective coordinates, and the analytical linear-mode analysis. The numerical and analytical results are then presented, covering the parameter-dependent bifurcation structures, Lyapunov exponents, Fourier spectra, magnetization trajectories, and FMR response. We then discuss these findings to elucidate the physical mechanisms underlying the non-linear dynamics and their relationship to the analytical collective modes, followed by the main conclusions.
\section{Model and Mathematical Formulation}

%\subsection{Physical model}

\begin{figure}
	\centering
	\includegraphics[width=0.5\textwidth]{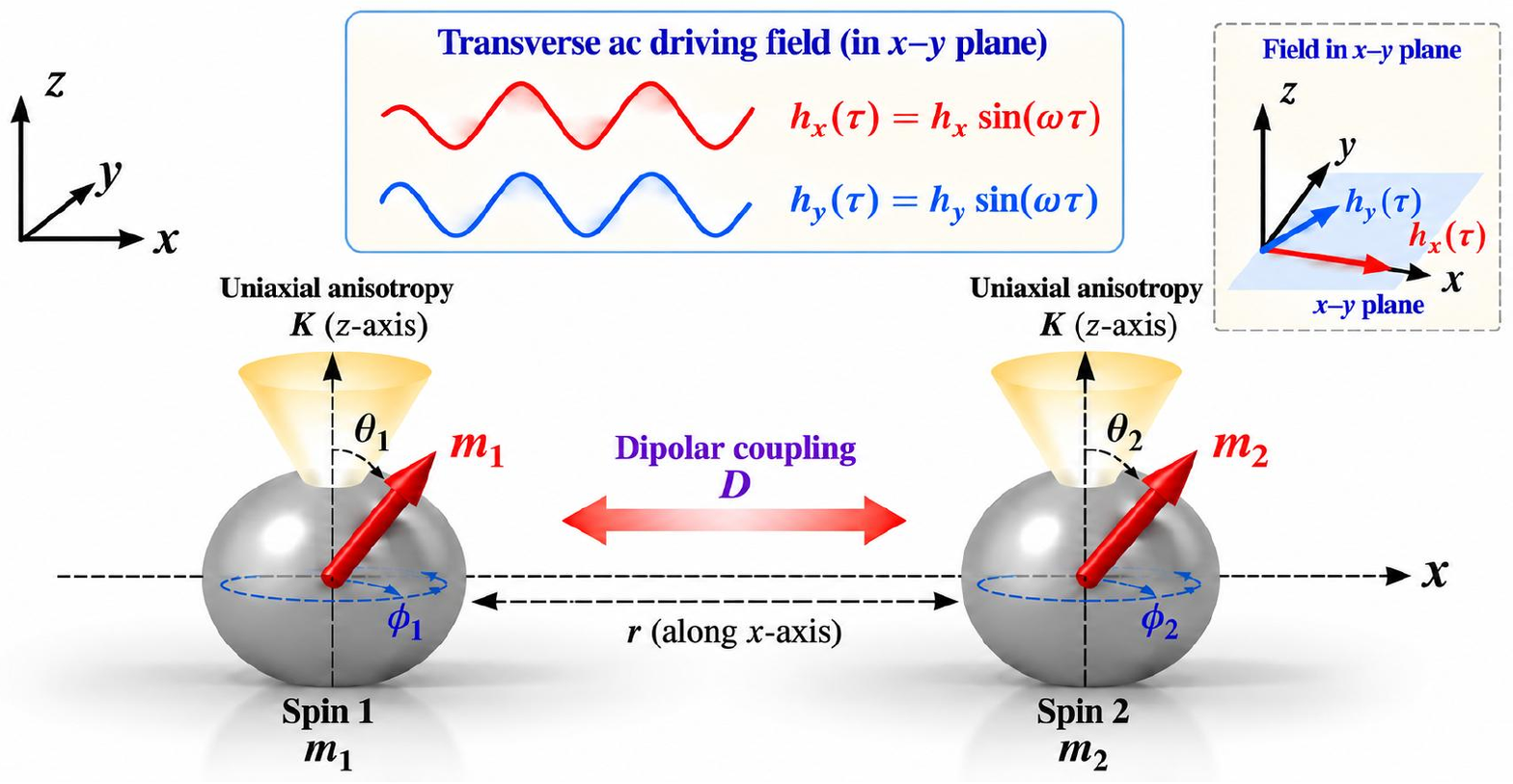}
           \vspace{-1cm}
	\caption{Schematic representation of two identical anisotropic magnetic spins separated along the $x$ axis and coupled through dipolar interaction of strength $D$. The easy axes are along $z$, while the transverse periodic excitation has components $h^{}_x\sin(\omega\tau)$ and $h^{}_y\sin(\omega\tau)$ in the $x$--$y$ plane.}
	\label{Figure1}
\end{figure}

We consider two identical magnetic moments
\begin{equation}
	\mathbf m^{}_i=(m^{}_{ix},m^{}_{iy},m^{}_{iz}),\qquad i=1,2
\end{equation}
with their centers separated along the $x$-axis. Both spins possess uniaxial anisotropy along $z$-axis. A static field $h_{\rm dc}$ is applied along $z$-axis, while a transverse periodic field acts in the $x$-$y$ plane [see schematic Fig.~(\ref{Figure1})] ,
\begin{equation}
	\mathbf h^{}_{\rm ac}(\tau)=
	(h^{}_x\hat{\mathbf x}+h^{}_y\hat{\mathbf y})\sin(\omega\tau)
\end{equation}
The spins are normalized, $|\mathbf m^{}_i|=1$, and interact through dipolar coupling.

%\subsection{Landau--Lifshitz--Gilbert dynamics}

The temporal evolution of the system can be modelled by the Landau- Lifshitz- Gilbert (LLG) equation~\cite{contreras2022,bragard2011,velez2020,velez2025}
\begin{equation}
	\frac{d\mathbf m^{}_i}{d\tau}
	=
	-\mathbf m^{}_i\times\mathbf h_i^{\rm eff}
	-\alpha\,\mathbf m^{}_i\times
	(\mathbf m^{}_i\times\mathbf h_i^{\rm eff}),
	\label{eq:llg}
\end{equation}
where $\alpha$ is the damping parameter. The above equation is in dimensionless form with $\mathbf {m}=\mathbf {M}/M^{}_s$; $M^{}_s$ being the saturation magnetization, $\mathbf h_i^{\rm eff}=\mathbf H_i^{\rm eff}/M^{}_s$; $\tau=t\gamma M^{}_s$; $\gamma=2.21\times 10^5 {\mathrm m A^{-1} s^{-1}}.$

Therefore, the effective field acting on the $i$-th spin is
\begin{equation}
	\mathbf h_i^{\rm eff}
	=
	(h^{}_x\hat{\mathbf x}+h^{}_y\hat{\mathbf y})\sin(\omega\tau)
	+K m^{}_{iz}\hat{\mathbf z}
	+h^{}_{\rm dc}\hat{\mathbf z}
	+\mathbf h_i^{\rm dip}
	\label{eq:heff}
\end{equation}
Here $K$ is the anisotropy strength (dimensionless), and $\mathbf h_i^{\rm dip}$ is the dipolar field acting on $i$-th spin defined as~\cite{urzagasti2015} 
%\begin{widetext} 
\begin{equation}
	\mathbf h_i^{\rm dip}=\frac{V}{4\pi}\sum_{j\ne i}
	\frac{3(\mathbf m_j\cdot\hat{\mathbf r}_{ij})\hat{\mathbf r}_{ij}
		-\mathbf m_j}{{r}_{ij}^{3}}=D\sum_{j\ne i}\big[
	 3(\mathbf m_j\cdot\hat{\mathbf r}_{ij})\hat{\mathbf r}_{ij}
		-\mathbf m_j\big]
	\label{eq:hdip}
\end{equation}
%\end{widetext}
Here $D=\frac{V}{4\pi r^3}$ is the dipolar field strength; it captures the long-range spin-spin interaction The effective field acting on spin 1 and spin 2 can therefore be expressed in components form for the system considered as
%\begin{widetext}
\begin{align}
	h_{1x}^{\rm eff}&=h^{}_x\sin(\omega\tau)+2Dm^{}_{2x},&\nonumber
	h_{1y}^{\rm eff}&=h^{}_y\sin(\omega\tau)-Dm^{}_{2y},&\\ \nonumber
	h_{1z}^{\rm eff}&=h^{}_{\rm dc}-Dm^{}_{2z}+Km^{}_{1z};& \nonumber
	h_{2x}^{\rm eff}&=h^{}_x\sin(\omega\tau)+2Dm^{}_{1x},&\\\nonumber
	h_{2y}^{\rm eff}&=h^{}_y\sin(\omega\tau)-Dm^{}_{1y},&
	h_{2z}^{\rm eff}&=h^{}_{\rm dc}-Dm^{}_{1z}+Km^{}_{2z}\\
\end{align}
%\end{widetext}
It is evident from above that for the chosen geometry, the dipolar-field strength differ between the spatial directions: the $x$ component contains $2D$, whereas the $y$ and $z$ components contain $-D$. Consequently, the interaction affects different magnetization components unequally. It is therefore quite obvious that the two-spin system provides a minimal model in which collective magnetic degrees of freedom and interaction-dependent instabilities can be studied. Since $h^{}_x$ and $h^{}_y$ multiply the same sinusoidal function, the applied field is linearly polarized along a fixed transverse direction, rather than circularly rotating.

%The Landau--Lifshitz--Gilbert (LLG) equations combine precession, damping, and periodic forcing. Precession generates nonlinear magnetization motion, damping dissipates energy, and the external field continuously drives the system away from equilibrium. These ingredients permit regular oscillations, complex nonlinear motion, and deterministic chaos without requiring stochastic fluctuations.

In order to avoid numerical artefacts~\cite{,velez2025,alvarez2000,perez2013classical}, it is suitable to
solve Eq.~(\ref{eq:llg}) in the Cartesian representation, namely,
\begin{widetext}
\begin{align}
	\frac{dm^{}_{1x}}{d\tau}={}&
	-h^{}_{dc}m^{}_{1y}
	+Dm^{}_{2z}m^{}_{1y}
	-Km^{}_{1z}m^{}_{1y}
	+m^{}_{1z}h^{}_y\sin(\omega\tau)
	-Dm^{}_{2y}m^{}_{1z} \nonumber\\
	&-\alpha\Big[
	m^{}_{1x}m^{}_{1y}h^{}_y\sin(\omega\tau)
	-Dm^{}_{2y}m^{}_{1x}m^{}_{1y}
	-m_{1y}^{2}h_x\sin(\omega\tau) -2Dm^{}_{2x}m_{1y}^{2}
	-m_{1z}^{2}h_x\sin(\omega\tau)\nonumber\\
	&\quad
	-2Dm^{}_{2x}m_{1z}^{2} +h^{}_{dc}m^{}_{1x}m^{}_{1z}
	-Dm^{}_{2z}m^{}_{1x}m^{}_{1z}
	+Km_{1z}^{2}m^{}_{1x}
	\Big],\label{eq:llg1x}\\
	\frac{dm^{}_{1y}}{d\tau}={}&
	-m^{}_{1z}h_x\sin(\omega\tau)
	-2Dm^{}_{2x}m^{}_{1z}
	+h^{}_{dc}m^{}_{1x}
	-Dm^{}_{2z}m^{}_{1x}
	+Km^{}_{1z}m^{}_{1x}\nonumber\\
	&-\alpha\Big[h^{}_{dc}m^{}_{1y}m^{}_{1z}-Dm^{}_{2z}m^{}_{1y}m^{}_{1z}+Km^{2}_{1z}m^{}_{1y}-m^{2}_{1z}h^{}_y\sin(\omega\tau)+Dm^{}_{2y}m^{2}_{1z}\nonumber\\
	&\quad -m^{2}_{1x}h^{}_y\sin(\omega\tau)+Dm_{2y}m^{2}_{1x}+m^{}_{1x}m^{}_{1y}h^{}_x\sin(\omega\tau)+2Dm^{}_{2x}m^{}_{1y}m^{}_{1x}\Big],\label{eq:llg1y}\\
	\frac{dm^{}_{1z}}{d\tau}={}&
	-m^{}_{1x}h^{}_y\sin(\omega\tau)
	+Dm^{}_{2y}m^{}_{1x}
	+m^{}_{1y}h^{}_x\sin(\omega\tau)
	+2Dm^{}_{2x}m^{}_{1y}\nonumber\\
	&-\alpha\Big[m^{}_{1x}m^{}_{1z}h^{}_x\sin(\omega\tau)+2Dm^{}_{1x}m^{}_{1z}m^{}_{2x}-h^{}_{dc}m^{2}_{1x}+Dm^{2}_{1x}m^{}_{2z}-Km^{2}_{1x}m^{}_{1z}\nonumber\\
	&\quad -h^{}_{dc}m^{2}_{1y}+Dm^{2}_{1y}m^{}_{2z}-Km^{2}_{1y}m^{}_{1z}+m^{}_{1y}m^{}_{1z}h^{}_y\sin(\omega\tau)-Dm^{}_{1y}m^{}_{1z}m^{}_{2y}
	\Big],\label{eq:llg1z}\\
	\end{align}
	\end{widetext}
	\begin{widetext}
	\begin{align}
	\frac{dm^{}_{2x}}{d\tau}={}&
	-h^{}_{dc}m^{}_{2y}+Dm^{}_{1z}m^{}_{2y}-Km^{}_{2z}m^{}_{2y}+m^{}_{2z}h^{}_y\sin(\omega\tau)-Dm^{}_{1y}m^{}_{2z}\nonumber\\
	&-\alpha\Big[h^{}_y\sin(\omega\tau)m^{}_{2x}m^{}_{2y}-Dm^{}_{1y}m^{}_{2x}m^{}_{2y}-h^{}_x\sin(\omega\tau)m^{2}_{2y}-2Dm^{}_{1x}m^{2}_{2y}-h^{}_x\sin(\omega\tau)m^{2}_{2z}\nonumber\\
	&\quad -2Dm^{}_{1x}m^{2}_{2z}+h^{}_{dc}m_{2x}m^{}_{2z}-Dm^{}_{1z}m^{}_{2x}m^{}_{2z}+Km^{}_{2x}m^{2}_{2z}\Big],\label{eq:llg2x}\\ 
	\frac{dm^{}_{2y}}{d\tau}={}&
	-h^{}_x\sin(\omega\tau) m^{}_{2z}-2Dm^{}_{1x}m^{}_{2z}+h^{}_{dc}m^{}_{2x}-Dm^{}_{1z}m^{}_{2x}+Km^{}_{2z}m^{}_{2x}\nonumber\\
	&-\alpha\Big[h^{}_{dc}m^{}_{2y}m^{}_{2z}-Dm^{}_{1z}m^{}_{2y}m^{}_{2z}+Km^{}_{2y}m^{2}_{2z}-h^{}_y\sin(\omega\tau)m^{2}_{2z}+Dm^{}_{1y}m^{2}_{2z}\nonumber\\
	&\quad -h^{}_y\sin(\omega\tau)m^{2}_{2x}+Dm^{}_{1y}m^{2}_{2x}+h^{}_x\sin(\omega\tau)m^{}_{2x}m^{}_{2y}+2Dm^{}_{1x}m^{}_{2x}m^{}_{2y}\Big],\label{eq:llg2y}\\
	\frac{dm^{}_{2z}}{d\tau}={}&
	-h^{}_y\sin(\omega\tau)m^{}_{2x}+Dm^{}_{1y}m^{}_{2x}+h^{}_x\sin(\omega\tau)m^{}_{2y}+2Dm^{}_{1x}m^{}_{2y}
	\nonumber\\
	&-\alpha\Big[ h^{}_x\sin(\omega\tau)m^{}_{2x}m^{}_{2y}+2Dm^{}_{1x}m^{}_{2x}m^{}_{2z}-h^{}_{dc}m^{2}_{2x}+Dm^{}_{1z}m^{2}_{2x}-Km^{2}_{2x}m^{}_{2z}\nonumber\\
	&\quad -h_{dc}m^{2}_{2y}+Dm^{}_{1z}m^{2}_{2y}-Km^{}_{2z}m^{2}_{2y}+h^{}_y\sin(\omega\tau)m_{2y}m^{}_{2z}-Dm_{1y}m_{2y}m^{}_{2z}\Big] \label{eq:llg2z}	\nonumber\\
\end{align}
\end{widetext}
\subsection{Collective coordinates}
We introduce
\begin{equation}
	\mathbf m^{}_S=\frac{\mathbf m^{}_1+\mathbf m^{}_2}{\sqrt2},
	\qquad
	\mathbf m^{}_A=\frac{\mathbf m^{}_1-\mathbf m^{}_2}{\sqrt2}
	\label{eq:collective}
\end{equation}
Therefore
%\begin{widetext}
\begin{equation}
	m^{}_{Sx}=\frac{m^{}_{1x}+m^{}_{2x}}{\sqrt2},
	%\qquad
	m^{}_{Sy}=\frac{m^{}_{1y}+m^{}_{2y}}{\sqrt2},
	%\qquad
	m^{}_{Sz}=\frac{m^{}_{1z}+m^{}_{2z}}{\sqrt2}
	\label{sym}
\end{equation}
%\end{widetext}
%\begin{widetext}
\begin{equation}
	m^{}_{Ax}=\frac{m^{}_{1x}-m^{}_{2x}}{\sqrt2},
	m^{}_{Ay}=\frac{m^{}_{1y}-m^{}_{2y}}{\sqrt2},
	m^{}_{Az}=\frac{m^{}_{1z}-m^{}_{2z}}{\sqrt2}
	\label{Anti}
\end{equation}
%\end{widetext}
The symmetric modes $m^{}_{Sx}$, $m^{}_{Sy}$, and $m^{}_{Sz}$ represent predominantly in-phase motion, whereas the antisymmetric modes $m^{}_{Ax}$, $m^{}_{Ay}$, and $m^{}_{Az}$ represent out-of-phase motion.

\subsection{Linear collective-mode analysis}

For the collinear reference state $\mathbf m^{}_1=\mathbf m^{}_2=\hat{\mathbf z}$, with the AC field absent, Eq.~(\ref{eq:llg1x})-Eq.~(\ref{eq:llg2z}) takes the following form

\begin{align}
	\frac{dm^{}_{1x}}{d\tau}&=-(h^{}_{dc}-D+K)m^{}_{1y}-Dm^{}_{2y} \label{fmrlg1x}\\
	\frac{dm^{}_{1y}}{d\tau}&=-2Dm^{}_{2x}+(h^{}_{dc}-D+K)m^{}_{1x}\label{fmrlg1y}\\
	0&=Dm^{}_{2y}m^{}_{1x}+2Dm^{}_{2x}m^{}_{1y}\nonumber\\
	\frac{dm^{}_{2x}}{d\tau}&=-(h^{}_{dc}-D+K)m^{}_{2y}-Dm^{}_{1y}\label{fmrlg2x}\\
	\frac{dm^{}_{2y}}{d\tau}&=-2Dm^{}_{1x}+(h^{}_{dc}-D+K)m^{}_{2x}\label{fmrlg2y}\\
	0&=Dm^{}_{1y}m^{}_{2x}+2Dm^{}_{1x}m^{}_{2y}\nonumber
\end{align}

Using Eq.~(\ref{sym}) and Eq.~(\ref{Anti}) in Eq.~(\ref{fmrlg1x})- Eq.~(\ref{fmrlg2y}) and with some algebraic manipulation, we obtain
 \begin{equation}
 	\ddot m^{}_{Sx}+\omega_S^2m^{}_{Sx}=0
 \end{equation}
 \begin{equation}
 	\ddot m^{}_{Ax}+\omega_A^2m^{}_{Ax}=0
 \end{equation}
 Where $\omega_S^2=(H^{}_0+D)(H^{}_0-2D)$, $\omega_A^2=(H^{}_0-D)(H^{}_0+2D)$, and 
 $H^{}_0=h^{}_{\rm dc}-D+K$

\begin{equation}
	\omega_S^2=(h^{}_{\rm dc}+K)(h^{}_{\rm dc}+K-3D)
	\label{omega_s}
\end{equation}

\begin{equation}
	\omega_A^2=(h^{}_{\rm dc}+K-2D)(h^{}_{\rm dc}+K+D)
		\label{omega_a}
\end{equation}
These expressions for $\omega^{}_S$ and $\omega^{}_A$ are the angular frequency of normal modes of symmetric and antisymmetric modes of the underlying system in the linear regime. It could therefore provide the linear reference for interpreting the numerical FMR spectra.

\subsection{Numerical procedure and dynamical diagnostics}

We integrate  Eq.~(\ref{eq:llg1x})-~(\ref{eq:llg2z}) using fourth-order Runge-Kutta method~\cite{botha2026landau,laroze2008dynamics} with
\begin{equation}
	\alpha=0.05,\qquad \Delta\tau=0.01,\qquad \tau_{\max}=1\times 10^4.
\end{equation}
For non-linear parameter scans, $K=1$, $h_{\rm dc}=0.10$, and $\omega=1$ are used unless otherwise specified. To eliminate transient effects, the initial $40\%$ of each simulated trajectory is discarded, and the subsequent $60\%$ is analyzed using Poincaré sampling, local maxima, Lyapunov exponents, and Fourier spectra.
For a periodically driven system,
\begin{equation}
	T=\frac{2\pi}{\omega},
\end{equation}
and the Poincaré section is obtained by sampling at equal phases of the drive. The Lyapunov exponents (LEs) evaluations consist of quantifying the divergence between two initially close trajectories of vector field. In general, for an effective N-dimensional dynamical system described by a set of equations $\frac{dX^{}_{i}}{d\tau}= F^{}_{i}(X,\tau)$, the $i$-th Lyapunov exponent is given by~\cite{skokos2009lyapunov,gallas2010,wolf1985}
\begin{equation}
	\lambda_i=
	\lim_{\tau\rightarrow\infty}
	\frac{1}{\tau}
	\ln\left[
	\frac{\|\delta\mathbf X^{}_i(\tau)\|}
	{\|\delta\mathbf X^{}_i(0)\|}
	\right],
	\label{eq:lyap}
\end{equation}
Where  $||\delta X^{\zeta}_{i}||$ is the distance between the trajectories of the $i$-th-component of the vector field at time $\zeta$. They can be ordered by decreasing amplitude: $\lambda^{}_{1}\ge \lambda^{}_{2}\ge...\ge \lambda^{}_{N}$. The first two exponents are the largest Lyapunov exponent (LLE) and the second-largest Lyapunov exponent (SLLE). To examine the sensitivity of the collective and relative spin responses, we additionally calculate the leading growth rates for perturbations initially oriented along symmetric and antisymmetric directions. These rates complement the full Lyapunov spectrum by characterizing the initial orientation of perturbation growth. For this, we additionally evaluate leading exponents associated with the symmetric and antisymmetric modes, $\lambda^{}_{S1}$ and $\lambda^{}_{A1}$.
{\color{black} The Lyapunov exponents are calculated using the Benettin–Gram–Schmidt algorithm by integrating the reference magnetization trajectory with its tangent perturbations. Since the two unit-length spins have four independent physical degrees of freedom, four Lyapunov exponents are evaluated from the six-component Cartesian representation, with perturbations projected onto the tangent planes of the spins. For the symmetric and antisymmetric analyses, we choose initial perturbation vectors \(\delta\mathbf m_1=\pm\delta\mathbf m_2\) and orthonormalize the corresponding two tangent vectors in each sector separately using the Gram–Schmidt procedure. To examine the stability of collective excitations, the Lyapunov exponents are evaluated separately using symmetric and antisymmetric initial perturbations. The symmetric perturbations satisfy \(\delta\mathbf m_1=\delta\mathbf m_2\), whereas the antisymmetric perturbations satisfy \(\delta\mathbf m_1=-\delta\mathbf m_2\). For each case, two orthonormal tangent vectors are evolved and periodically orthonormalized using the Gram–Schmidt procedure~\cite{skokos2009lyapunov,wiesel1993continuous,bosetti2017transient}. The exponents are obtained from the accumulated logarithmic stretching factors, averaged over the measurement interval. The simulations use \(\Delta\tau=0.01\), discard the transient up to \(\tau=4000\), and continue to \(\tau=10000\), with Gram–Schmidt orthonormalization every \(0.1\) time units. A positive largest Lyapunov exponent, supported by the Poincaré sections and magnetization spectra, identifies chaotic dynamics.}

%\subsection{Dipolar coupling and nonlinear dynamical complexity}

\section{Simulation and Analytical Results}
{\color{black} We first consider dipolar coupling as the principal control parameter to investigate the Poincaré-sampled bifurcation diagrams and corresponding Lyapunov exponents. Fig.~(\ref{Figure2}) presents the evolution of the magnetization dynamics and Lyapunov exponents as functions of the dipolar coupling strength $D$ for two driving amplitudes, $h^{}_x=h^{}_y=1.0$ [Figs. 2(a)–2(e)] and $h^{}_x=h^{}_y=2.0$ [Figs. 2(f)–2(j)].  The bifurcation diagrams are constructed from 
$m^{}_{1x}$, $m^{}_{Sx}$ and $m^{}_{Ax}$, while the lower panels display $\lambda^{}_1$, $\lambda^{}_2$, $\lambda^{}_{S1}$, and $\lambda^{}_{A1}$.
At relatively weak coupling, the stroboscopic responses contain comparatively organized branches [see Fig.~\ref{Figure2}(a)-(c)]. As $D$ increases, these branches change, split, and broaden into dense regions, with narrower windows of organized response reappearing at some coupling values. This behaviour indicates repeated reorganizations of the long-time dynamics rather than a monotonic increase in chaos. Possible mechanisms include bifurcations of periodic orbits, transitions between coexisting attractors, and periodic windows within chaotic regimes. The collective coordinates further reveal the role of relative spin dynamics. While $m^{}_{Sx}$ characterizes the combined transverse response, $m^{}_{Ax}$ measures the difference between the transverse magnetizations of the two spins. Their distinct coupling-dependent structures suggest that the non-linear response involves both collective and relative degrees of freedom.

\begin{figure*}
	\centering
	\includegraphics[width=0.8\textwidth]{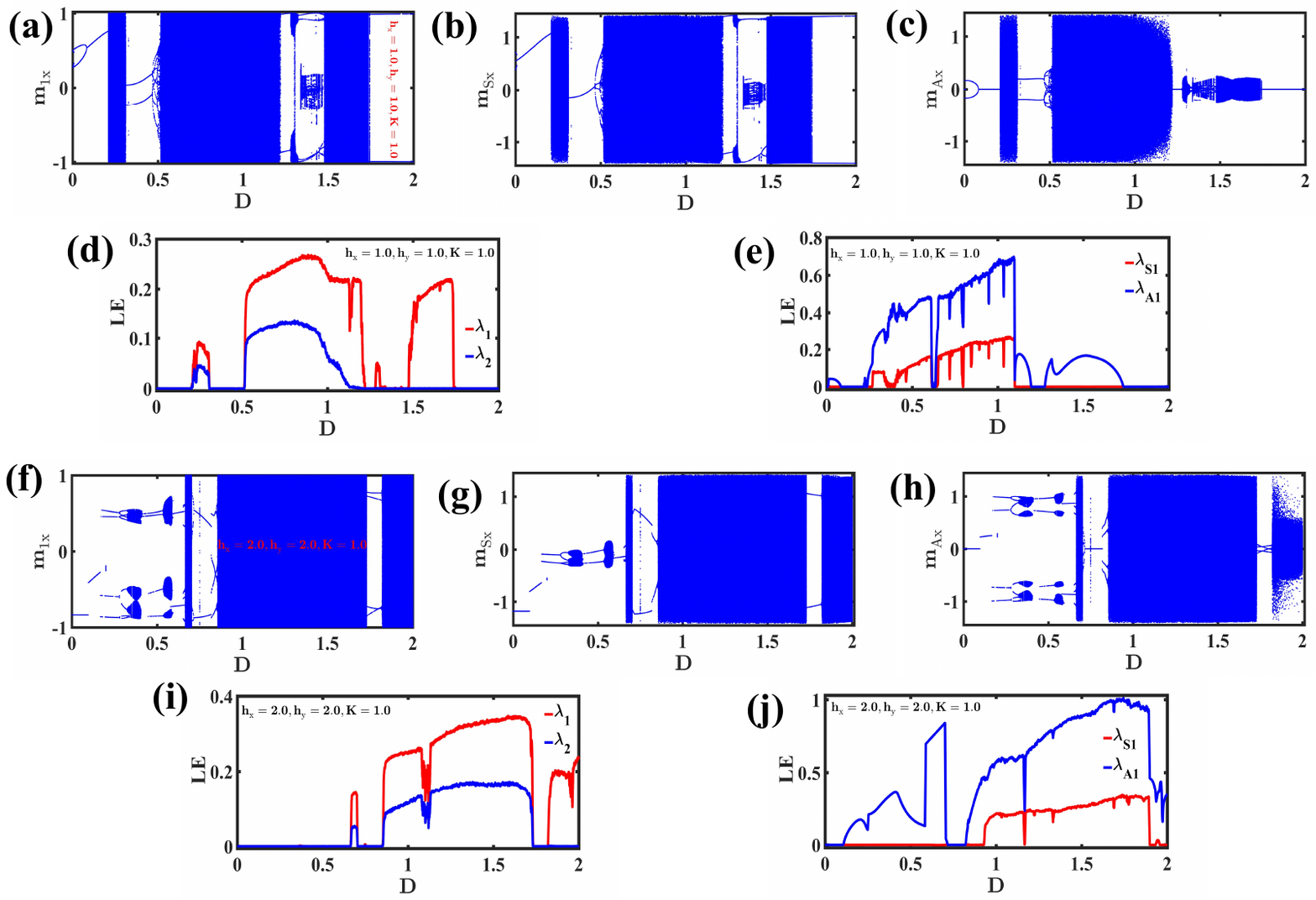}
	\vspace{-1cm}
	\caption{Poincaré-sampled bifurcation diagrams and corresponding Lyapunov-exponent analysis as a function of dipolar coupling $D$ for representative transverse driving amplitudes. The bifurcation diagrams are constructed from $m^{}_{1x}$, $m^{}_{Sx}$ and $m^{}_{Ax}$, while the lower panels show $\lambda^{}_1$, $\lambda^{}_2$, $\lambda^{}_{S1}$, and $\lambda^{}_{A1}$.}
	\label{Figure2}
\end{figure*}

The corresponding Lyapunov exponents in Fig~.\ref{Figure2}(d) and Fig.~\ref{Figure2}(i) provide a quantitative measure of these changes. The largest Lyapunov exponent $\lambda^{}_{1}$ becomes positive over several intervals of $D$, supporting the occurrence of deterministic chaos in these regions. Conversely, intervals in which the leading exponent approaches zero coincide with a reduction in exponential sensitivity, consistent with more regular or weakly unstable dynamics. The non-monotonic variation of the exponents demonstrates that the dipolar interaction reorganizes the stability of the driven spin system, producing alternating regions of enhanced and suppressed instability. In addition, it is also evident that second largest Lyapunov exponent $\lambda^{}_{2}$ is also positive in unison with $\lambda^{}_{1}$ for quite large range of $D$ indicating behaviour consistent with hyperchaos. This distinct bifurcation structure corresponding to symmetric and antisymmetric mode is reinforced by the perturbation-growth results in Fig.~\ref{Figure2}(i)  and Fig.~\ref{Figure2}(j), where the leading growth rate associated with antisymmetric initial perturbations exceeds that associated with symmetric perturbations over substantial coupling intervals. Thus, perturbations that initially differentiate the spin trajectories can be more strongly amplified than perturbations that initially displace them together. 
%These curves characterize the growth of the specified initial perturbations and should not be interpreted as independent mode-specific Lyapunov spectra unless the corresponding symmetry sectors remain invariant under the dynamics.

 %The variation of $\lambda^{}_{S1}$ and $\lambda^{}_{A1}$ also support the bifurcation characteristics with $\lambda^{}_{A1}$ dominates over $\lambda^{}_{S1}$.     

%The collective coordinates further reveal the role of relative spin dynamics. While $m^{}_{Sx}$ characterizes the combined transverse response, $m^{}_{Ax}$ measures the difference between the transverse magnetizations of the two spins. Their distinct coupling-dependent structures suggest that the non-linear response involves both collective and relative degrees of freedom. This distinction is reinforced by the perturbation-growth results in Fig.~\ref{Figure2}(i)  and Fig.~\ref{Figure2}(j), where the leading growth rate associated with antisymmetric initial perturbations exceeds that associated with symmetric perturbations over substantial coupling intervals. Thus, perturbations that initially differentiate the spin trajectories can be more strongly amplified than perturbations that initially displace them together. These curves characterize the growth of the specified initial perturbations and should not be interpreted as independent mode-specific Lyapunov spectra unless the corresponding symmetry sectors remain invariant under the dynamics.

Comparison of the two driving amplitudes ($h^{}_x=h^{}_y=1.0$ and $h^{}_x=h^{}_y=2.0$) reveals that stronger excitation changes the coupling intervals over which complex magnetization responses and positive Lyapunov exponents occur. The stronger drive therefore modifies the organization of the instability landscape rather than producing a uniform increase in chaos at all coupling strengths. Taken together, the magnetization maps and Lyapunov analysis demonstrate a non-trivial interplay between periodic excitation and dipolar coupling, with the stability of the driven two-spin system depending sensitively on both parameters.

%It establishes the nonmonotonic coupling dependence of the nonlinear response and links complex-response regions to positive Lyapunov exponents. The mode-oriented results suggest unequal sensitivity of symmetric and antisymmetric initial perturbations.

%\subsection{Independent characterization using local maxima}

%The local maxima of the magnetization response provide an independent characterization of the same dynamical structures. For a simple periodic response, the maxima form a small number of well-defined branches. As the dynamics become more complicated, these branches split and form increasingly dense distributions. The principal structures are consistent with those obtained from Poincare sampling.

\begin{figure*}
	\centering
	\includegraphics[width=0.8\textwidth]{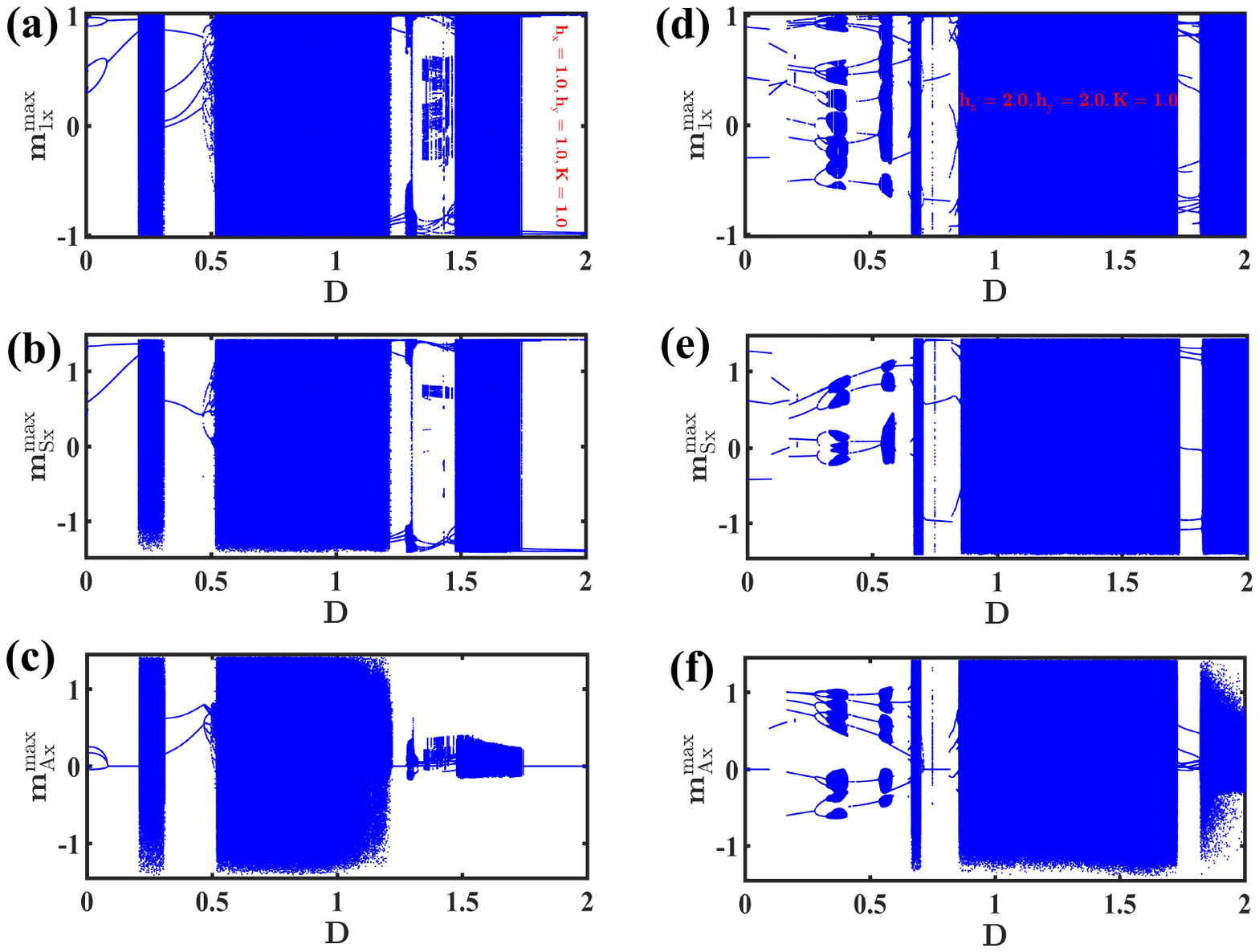}
	\vspace{-0.5cm}
	\caption{Bifurcation diagrams based on local maxima of the magnetization response as a function of dipolar coupling $D$ for representative driving amplitudes.}
	\label{Figure3}
\end{figure*}

In Fig.~(\ref{Figure3}), we study the bifurcation characteristics obtained from the local maxima of the magnetization response as $D$ varies, for the same driving amplitudes considered in Fig.~(\ref{Figure2}). Unlike Poincaré sampling, which records the magnetization at a fixed phase of the external drive, the local-maxima analysis follows the extrema of the continuous-time trajectory. For a periodic orbit, a magnetization component generally produces a finite set of recurring maxima. Splitting of these branches or the appearance of additional branches indicates a change in the periodic response and may reflect a change in orbit period or stability. In more complex regimes, the maxima become more broadly distributed because successive oscillations no longer repeat in the same way. The local-maxima diagrams therefore provide an independent check of the structures identified by Poincar\'e sampling. Their agreement supports the interpretation that the response reorganizes over corresponding intervals of dipolar coupling. Nevertheless, branch proliferation alone does not establish chaos, since high-period and quasiperiodic motion can also produce complicated patterns; the Lyapunov exponents provide the complementary stability diagnostic.

%\subsection{Control by transverse excitation}

%Figure.~(\ref{Figure4}) presents the Poincare bifurcation diagrams and Lyapunov exponents as a function of $h_x$ for representative $D$, with $h_y=1$. Increasing $h_x$ changes the energy supplied to the system and produces substantial changes in the dynamical attractor. The corresponding Lyapunov exponents demonstrate that excitation amplitude can destabilize otherwise regular states. The different behavior of $\lambda_{S1}$ and $\lambda_{A1}$ indicates that the drive modifies the stability of the collective modes rather than simply increasing the overall oscillation amplitude.

\begin{figure*}
	\centering
	\includegraphics[width=0.8\textwidth]{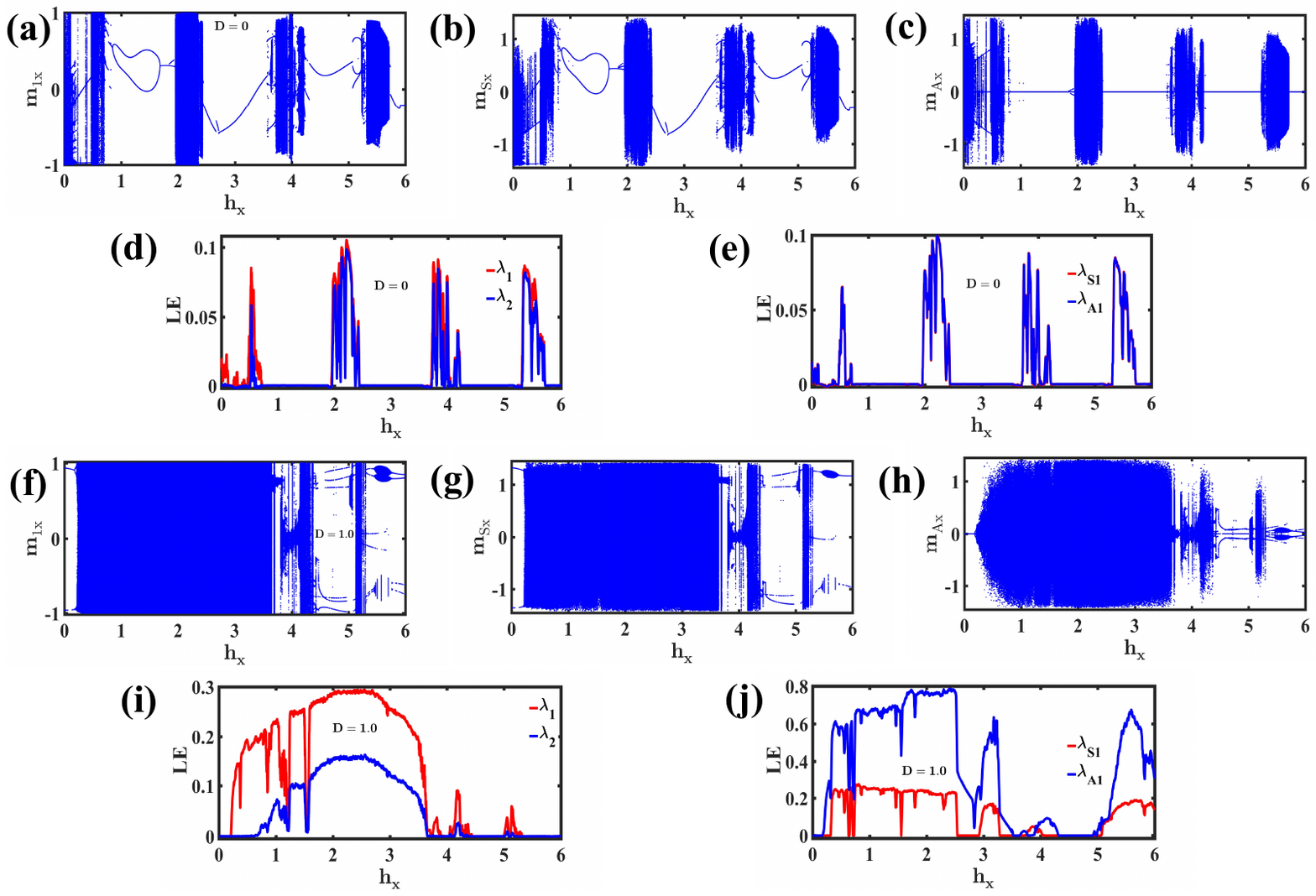}
	\vspace{-1cm}
	\caption{Poincaré-sampled bifurcation diagrams and corresponding Lyapunov-exponent analysis as a function of transverse driving amplitude $h^{}_x$ for representative dipolar coupling strength $D$ keeping $h^{}_y$ fixed. The bifurcation diagrams are constructed from $m^{}_{1x}$, $m^{}_{Sx}$ and $m^{}_{Ax}$, while the lower panels show $\lambda^{}_1$, $\lambda^{}_2$, $\lambda^{}_{S1}$, and $\lambda^{}_{A1}$.}
	\label{Figure4}
\end{figure*}

In Fig.~(\ref{Figure4}), we investigate the bifurcation response and Lyapunov exponents as functions of $h_x$, keeping $h_y=1.0$ fixed for representative dipolar coupling strengths, $D=0$ and $D=1$. The bifurcation diagrams show a non-monotonic sequence of organized and complex responses as $h_x$ increases, rather than a uniform progression from regular to chaotic motion. The Lyapunov exponents support this interpretation: intervals with positive leading exponents indicate exponential sensitivity to initial conditions, whereas values close to zero are consistent with more regular or weakly unstable dynamics. The symmetric- and antisymmetric-perturbation results show related but non-identical variations. At $D=1$, the diagrams indicate a broader range of complex response than at $D=0$, while the simultaneous positivity of the two largest Lyapunov exponents is consistent with hyper-chaotic dynamics. The difference between the symmetric and antisymmetric perturbation-growth curves indicates that changing $h_x$ affects perturbations of the combined and relative spin motion unequally. This behaviour can be understood due to modification of the time-dependent torque acting on the spins and hence changes both the amplitude and non-linear structure of their motion due to $h^{}_x$.

%Changing $h_x$ modifies the time-dependent torque acting on the spins and hence changes both the amplitude and nonlinear structure of their motion.

%The bifurcation diagrams show how the recurring magnetization values evolve with $h_x$. The Lyapunov exponents identify the parameter intervals associated with exponential sensitivity. Positive largest exponents support drive-induced chaos, whereas near-zero values are consistent with more regular or weakly unstable responses. Differences between the symmetric and antisymmetric perturbation-growth curves indicate that the drive can affect perturbations of the combined and relative spin motion unequally.

%The transverse excitation amplitude is an independent control parameter for nonlinear stability. The relevant effect is not merely an increase in oscillation amplitude, but a change in the parameter-dependent organization of dynamical regimes.

%The complementary dependence on $h_y$, with $h_x=1$, is shown in Fig.~(\ref{Figure5}). The response to the two transverse components is not identical because the dipolar field is anisotropic for the chosen geometry. Thus, $h_x$ and $h_y$ constitute physically distinct control parameters.

\begin{figure*}
	\centering
	\includegraphics[width=0.8\textwidth]{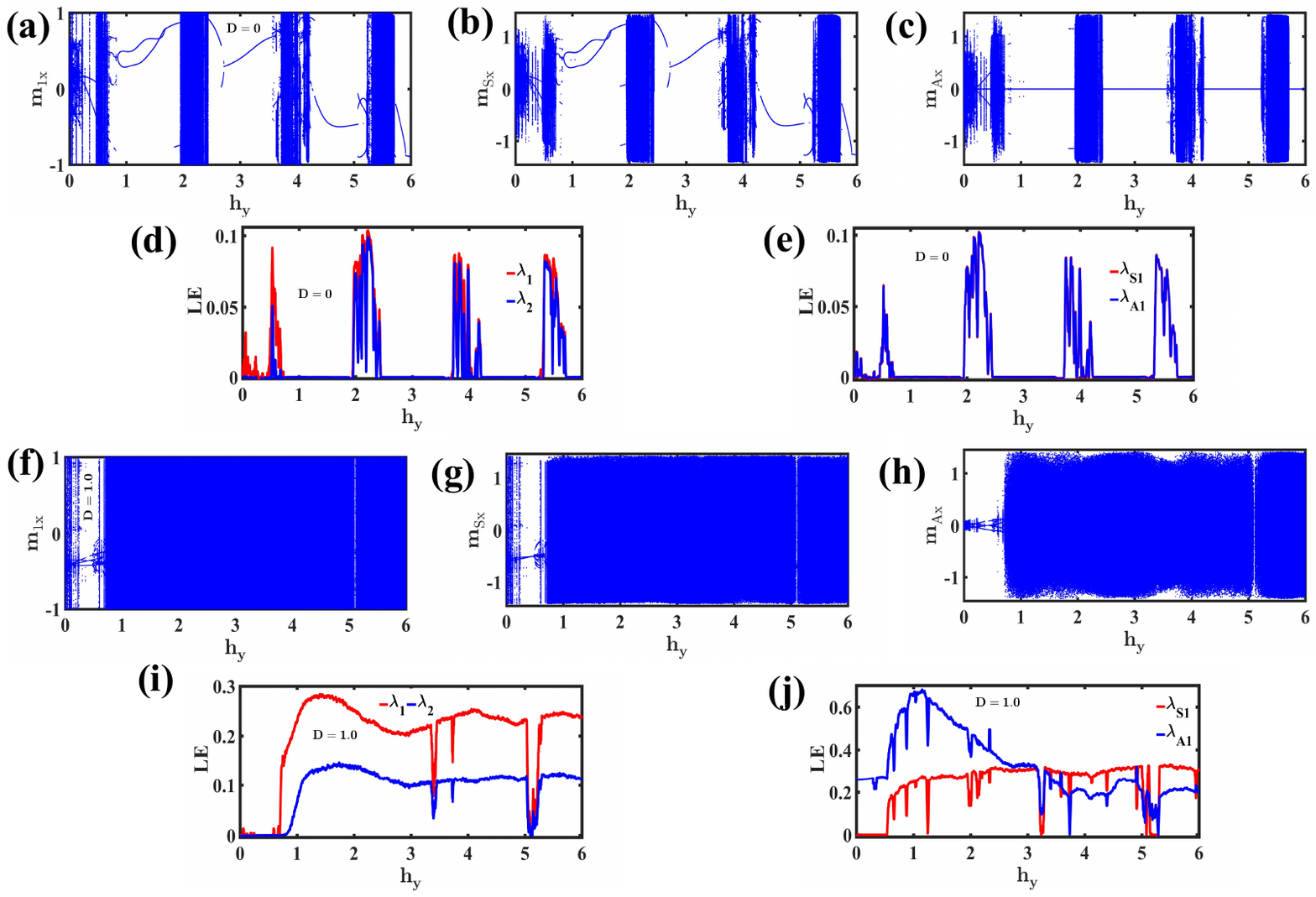}
	\vspace{-1cm}
	\caption{Poincaré-sampled bifurcation diagrams and corresponding Lyapunov-exponent analysis as a function of transverse driving amplitude $h^{}_y$ for representative dipolar coupling strength $D$ keeping $h^{}_x$ fixed. The bifurcation diagrams are constructed from $m^{}_{1x}$, $m^{}_{Sx}$ and $m^{}_{Ax}$, while the lower panels show $\lambda^{}_1$, $\lambda^{}_2$, $\lambda^{}_{S1}$, and $\lambda^{}_{A1}$.}
	\label{Figure5}
\end{figure*}

In Fig.~(\ref{Figure5}), we examine the Poincar\'e-sampled bifurcation response and Lyapunov exponents as functions of $h_y$, keeping $h_x=1.0$ fixed for $D=0$ and $D=1$. As for the $h_x$ sweep, the response varies non-monotonically with drive amplitude, and the Lyapunov exponent changes between positive and near-zero values. For $D=1$, the response is periodic or quasiperiodic at smaller $h_y$ in the plotted range, whereas for $h_y>0.5$ the diagram indicates a more persistent chaotic response. This behaviour differs from the $h_x$ dependence in Fig.~(\ref{Figure4}), demonstrating that the two drive components are not interchangeable. A small number of recurring branches is consistent with a regular periodic response, while branch proliferation and broader distributions indicate more complicated long-time dynamics; the Lyapunov exponents help distinguish chaotic regimes from complex but non-chaotic motion. Although $h_x$ and $h_y$ multiply the same sinusoidal time dependence, they act along different spatial directions. Because the dipolar field has different coefficients along $x$ and $y$, changing these amplitudes modifies the effective torque and the relative-spin response differently. Thus, $h_x$ and $h_y$ are distinct control parameters for the chosen geometry.

%\subsection{Effect of magnetic anisotropy}

%The effect of the uniaxial anisotropy is examined by varying $K$ at fixed $h_x=h_y=1$. The corresponding bifurcation diagrams and Lyapunov exponents are presented in Fig.~(\ref{Figure6}).Anisotropy determines the restoring torque against deviations from the easy axis and modifies the characteristic precessional dynamics. Its influence is explicit in the analytical collective-mode frequencies, where $K$ changes the frequency scales against which the periodic excitation acts. The nonlinear response therefore depends on the relative balance between anisotropy, dipolar interaction, and driving.

\begin{figure*}
	\centering
	\includegraphics[width=0.8\textwidth]{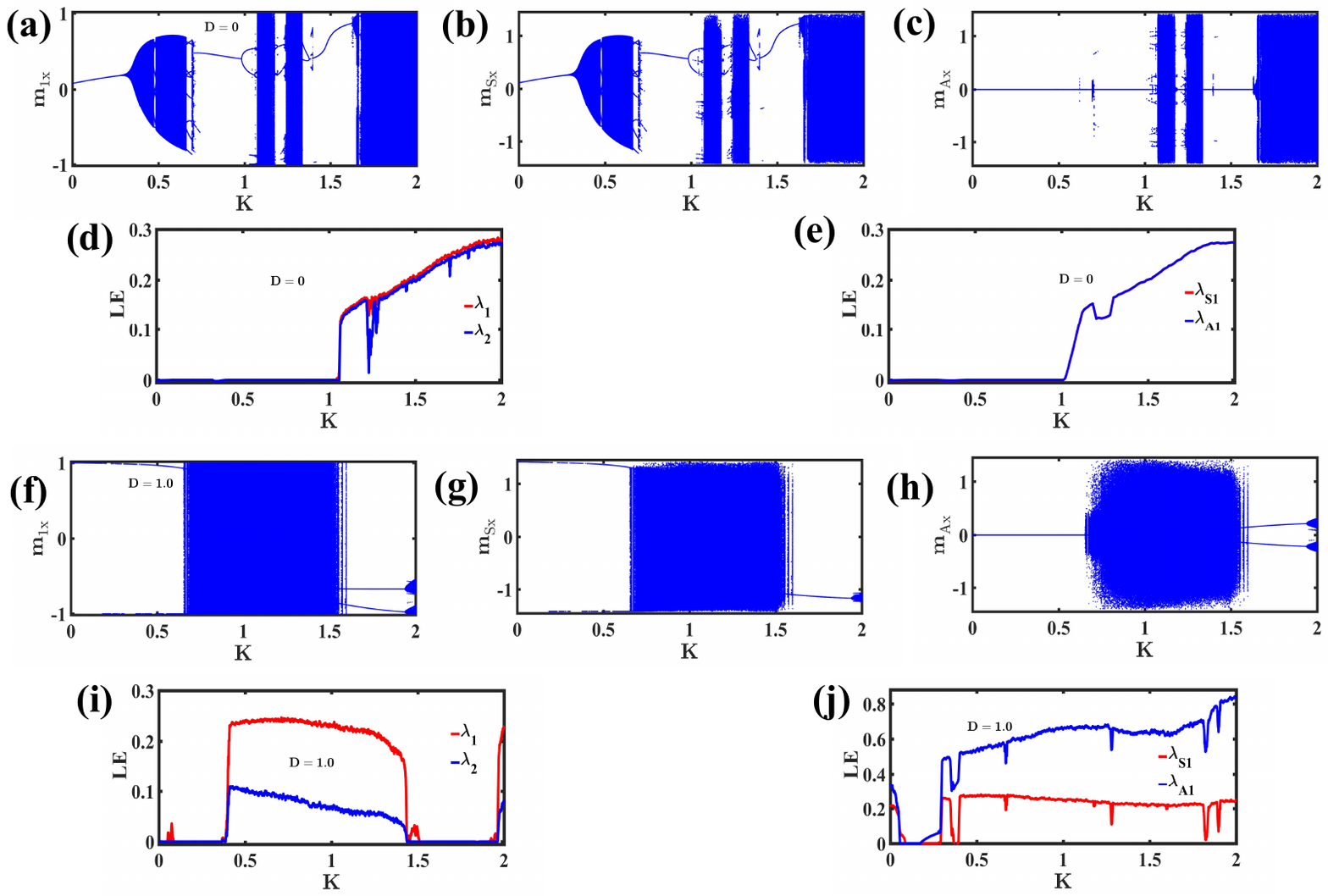}
	\vspace{-1cm}
	\caption{Poincaré-sampled bifurcation diagrams and corresponding Lyapunov exponents as a function of uniaxial anisotropy strength $K$ for representative values of $D$, with $h^{}_x=h^{}_y=1$. The bifurcation diagrams are constructed from $m^{}_{1x}$, $m^{}_{Sx}$ and $m^{}_{Ax}$, while the lower panels show $\lambda^{}_1$, $\lambda^{}_2$, $\lambda^{}_{S1}$, and $\lambda^{}_{A1}$.}
	\label{Figure6}
\end{figure*}

In Fig.~(\ref{Figure6}), we investigate the dependence of the bifurcation structures and Lyapunov exponents on the anisotropy strength $K$, with $h_x=h_y=1.0$. For uncoupled spins ($D=0$), periodic, quasiperiodic, and chaotic responses occur non-monotonically as $K$ varies. For coupled spins ($D=1$), chaotic states persist over a broader range of the plotted anisotropy values. The Lyapunov-exponent variation supports the changes identified in the bifurcation diagrams, and related structures are also visible in the symmetric- and antisymmetric-coordinate analyses. As in this model, the uniaxial anisotropy is directed along the $z$ axis and contributes a field proportional to $K m_{iz}$; consequently, it changes the restoring torque as a spin moves away from that axis. Varying $K$ therefore changes the balance between anisotropy, dipolar coupling, and periodic forcing. The resulting change in the precessional response can reorganize the non-linear trajectories and their stability. The contrasting behaviour at $D=0$ and $D=1$ shows that the influence of anisotropy cannot be considered independently of the interaction between the spins.
%Two-dimensional maps of $\lambda_1$, $\lambda_2$, $\lambda_{S1}$, and $\lambda_{A1}$ are used to characterize the multidimensional parameter space. Figure.~(\ref{Figure7}) presents maps in the $D-h_x$ and $D-h_y$ planes. The largest-Lyapunov-exponent maps reveal structured regions of instability rather than isolated parameter points. Comparison with the collective-mode maps demonstrates that different portions of the unstable parameter space can be associated with different collective sectors.

\begin{figure*}
	\centering
	\includegraphics[width=0.77\textwidth]{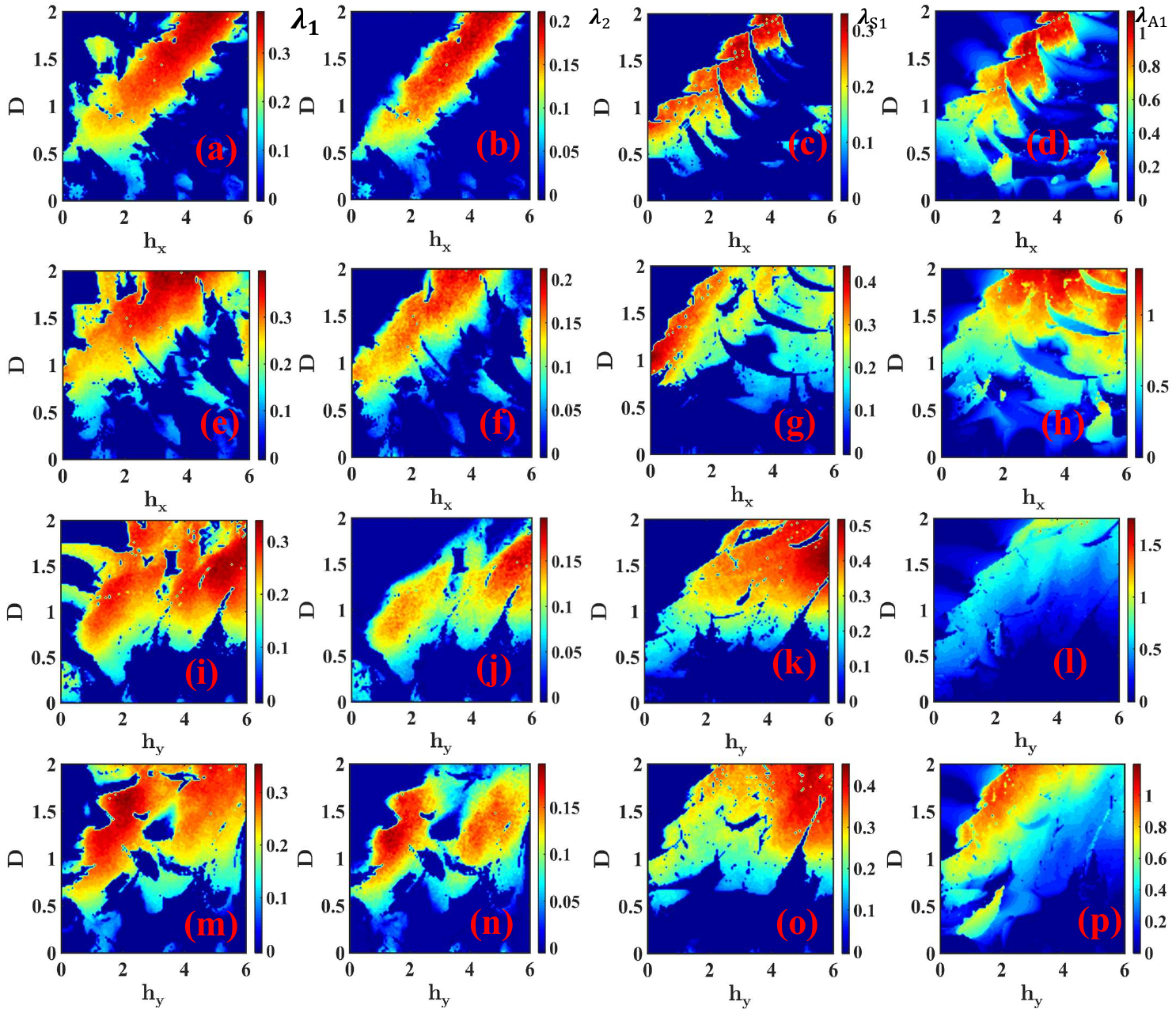}
	\caption{Two-dimensional maps of the largest and second-largest Lyapunov exponents, together with the leading symmetric- and antisymmetric-mode exponents, in the $D-h^{}_x$ and $D-h^{}_y$ parameter planes.}
	\label{Figure7}
\end{figure*}

In Fig.~(\ref{Figure7}), we show two-dimensional maps of $\lambda_1$, $\lambda_2$, $\lambda_{S1}$, and $\lambda_{A1}$ in the $D$--$h_x$ and $D$--$h_y$ parameter planes. Regions with positive leading Lyapunov exponents identify parameter ranges with exponential perturbation growth and are candidates for chaotic dynamics. The curved boundaries and bands show that the drive amplitude at which the stability changes depends on the dipolar coupling. In substantial regions, $\lambda_1$ and $\lambda_2$ are simultaneously positive, consistent with hyper-chaotic dynamics. The symmetric- and antisymmetric-perturbation maps provide a complementary comparison of growth rates for perturbations initially directed along the combined and relative spin coordinates. Their differing structures indicate that these perturbation classes respond differently across parameter space. Overall, the maps show that stability is controlled by the combined values of coupling and drive amplitude, rather than by either parameter in isolation.

\begin{figure*}
	\centering
	\includegraphics[width=0.8\textwidth]{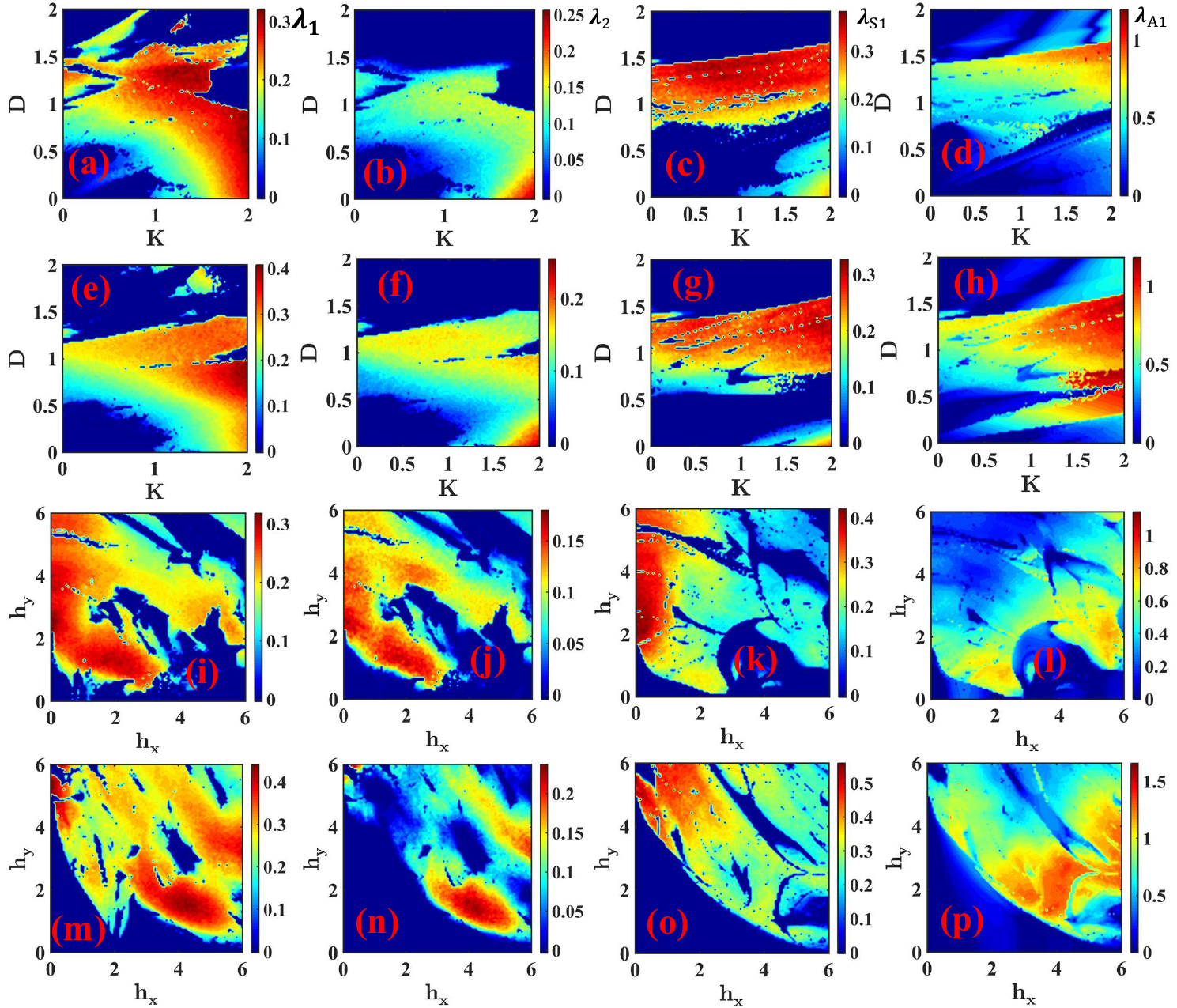}
	\caption{Two-dimensional maps of $\lambda^{}_1$, $\lambda^{}_2$, $\lambda^{}_{S1}$, and $\lambda^{}_{A1}$ in the $D-K$ and $h^{}_x-h^{}_y$ parameter planes.}
	\label{Figure8}
\end{figure*}

In Fig.~(\ref{Figure8}), we present Lyapunov maps in the $D$--$K$ and $h_x$--$h_y$ planes. The $D$--$K$ plane combines the dipolar interaction, which couples the spins anisotropically, with the uniaxial restoring torque. Their combined effect changes the effective fields and can alter both the characteristic precessional timescales and the stability of the driven trajectories. In the $h_x$--$h_y$ plane, the amplitudes of the two in-phase transverse field components are varied together. This changes the magnitude and direction of the fixed transverse drive. Since the dipolar interaction distinguishes the $x$ and $y$ directions, equal changes in $h_x$ and $h_y$ need not produce equivalent responses. The maps thus show how the computed growth rates depend on combinations of the interaction, anisotropy, and drive components. The contours identify regions of differing stability, but their appearance alone is insufficient to assign a specific bifurcation mechanism.

\begin{figure*}
	\centering
	\includegraphics[width=0.8\textwidth]{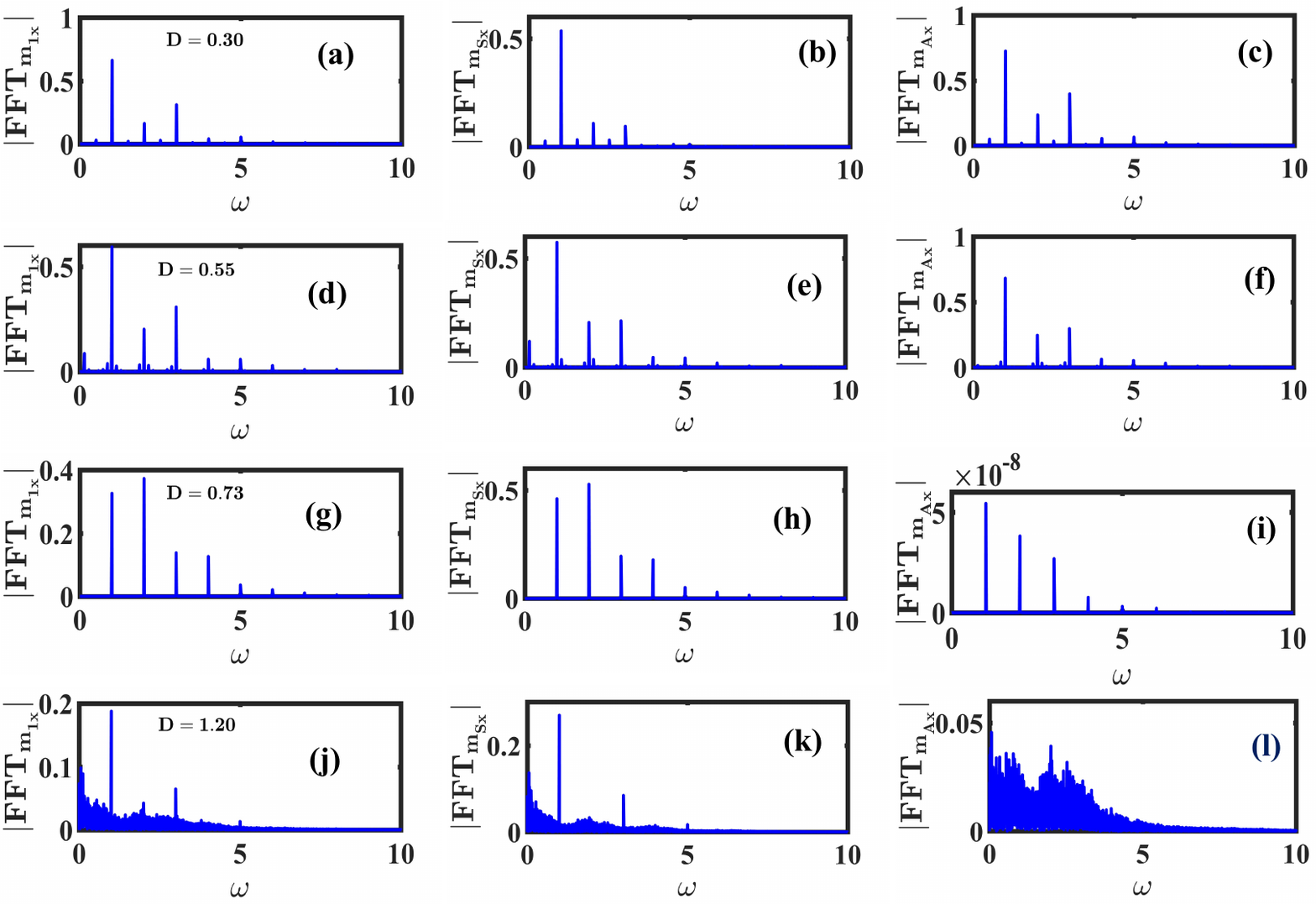}
	\vspace{-1cm}
	\caption{Fourier spectra of the magnetization dynamics for $D=0.30$, $0.55$, $0.73$, and $1.20$. For each value of $D$, spectra of the first-spin magnetization, symmetric mode, and antisymmetric mode are shown.}
	\label{Figure9}
\end{figure*}

In Fig.~(\ref{Figure9}), we present Fourier spectra for $D=0.30$, $0.55$, $0.73$, and $1.20$ for the first-spin, symmetric, and antisymmetric magnetization components. The spectra identify the frequencies contributing to the response and show how the spectral weight is distributed among them. At lower and intermediate coupling, distinct peaks and their harmonics are consistent with coherent non-linear oscillations. Harmonics can arise from the non-linear LLG dynamics even though the applied field has a single sinusoidal frequency. As $D$ changes, the relative peak strengths and the additional spectral components emerge, indicating that the interaction modifies the temporal structure of the magnetization response. The symmetric spectrum describes the combined spin response, whereas the antisymmetric spectrum emphasizes differences between the two spins; comparing them therefore helps identify which collective coordinate contributes most strongly at a given frequency. In particular, the very small antisymmetric spectral scale in some panels should be read as a small response amplitude, not as a divergence. The Fourier spectra complement the bifurcation diagrams and Lyapunov exponents: the spectra characterize frequency content, while the latter diagnostics provide information about the organization and stability of the dynamics.

%\subsection{Phase-space evolution}

%Figure.~(\ref{Figure10}) shows three-dimensional magnetization trajectories for $D=0.30$, $0.55$, $0.73$, and $1.20$. At lower dipolar coupling, the trajectories remain comparatively organized and occupy restricted regions of phase space. Increasing $D$ produces progressively more complicated trajectory structures. The collective representation reveals how this complexity is distributed between the symmetric and antisymmetric sectors.

\begin{figure*}
	\centering
	\includegraphics[width=0.8\textwidth]{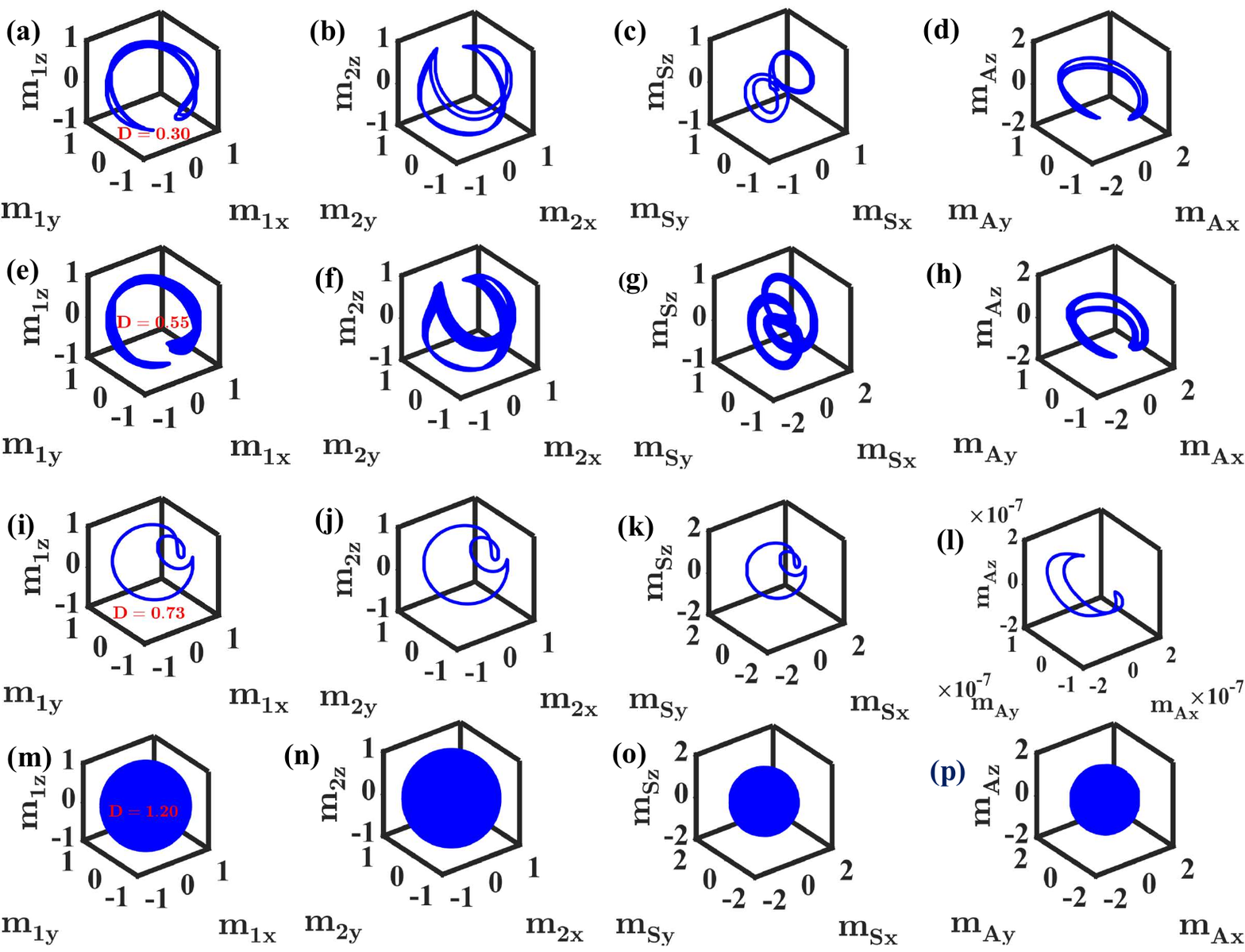}
	\caption{Three-dimensional magnetization trajectories for $D=0.30$, $0.55$, $0.73$, and $1.20$. For each value of $D$, trajectories are shown for the individual spins and for the symmetric and antisymmetric collective modes.}
	\label{Figure10}
\end{figure*}

In Fig.~(\ref{Figure10}), we show three-dimensional trajectories of the individual spins and the symmetric and antisymmetric collective coordinates for $D=0.30$, $0.55$, $0.73$, and $1.20$. These trajectories provide a geometrical view of the magnetization motion in state space. Relatively restricted loops or organized curves at lower coupling are consistent with periodic or quasiperiodic responses. As $D$ changes, the trajectories acquire different shapes and can occupy broader or denser regions, reflecting changes in the accessible motion as the dipolar fields and collective response are modified. The individual-spin trajectories show the motion of each moment, while the symmetric and antisymmetric coordinates separate the combined response from the relative motion of the two spins. Their comparison therefore helps indicate whether the observed geometrical complexity is expressed mainly in the collective or relative coordinates. However, a visually dense trajectory is not, by itself, proof of chaos; it should be interpreted together with the Poincar\'e diagrams (Bifurcation diagram) and Lyapunov exponents.

%\subsection{Ferromagnetic resonance and collective modes}

%The nonlinear calculations are performed at fixed driving frequency. To determine the underlying linear collective modes, we separately calculate the frequency-dependent FMR response. Figure.~(\ref{Figure11}) presents the FMR spectra of the first-spin magnetization and the symmetric and antisymmetric collective modes for representative $D$. The resonances shift with dipolar coupling and their amplitudes are redistributed between the collective channels.

\begin{figure*}[!htb]
	\centering
	\includegraphics[width=0.6\textwidth]{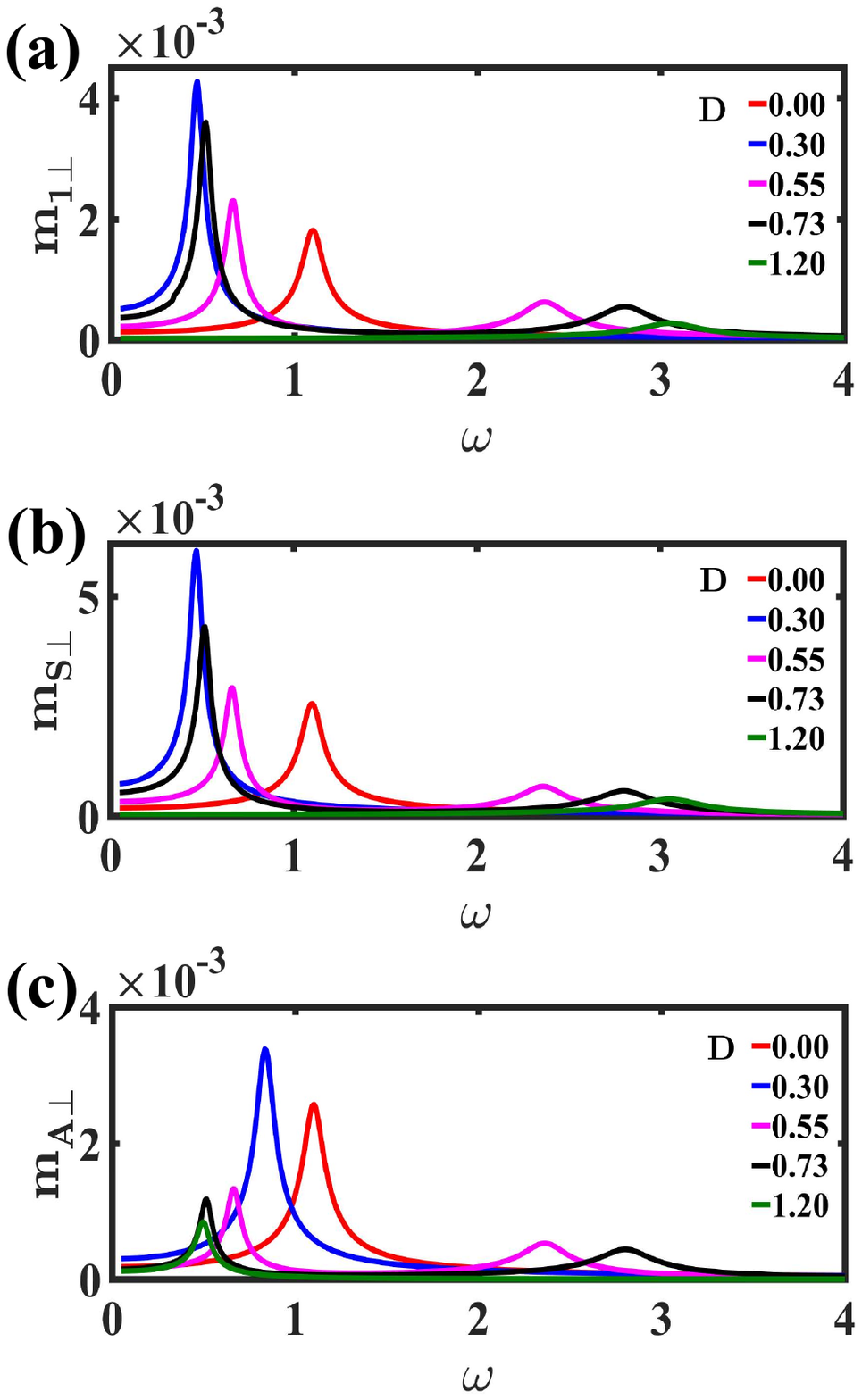}
	\vspace{-0.5cm}
	\caption{Ferromagnetic-resonance response as a function of driving angular frequency for representative values of $D$. The responses of the first-spin magnetization, symmetric collective mode, and antisymmetric collective mode are shown separately.}
	\label{Figure11}
\end{figure*}

To examine the underlying frequency-dependent response, we separately calculate the FMR spectra. Fig.~(\ref{Figure11}) shows the responses for $D=0$, $0.30$, $0.55$, $0.73$, and $1.20$. In contrast to the fixed-frequency non-linear scans, the excitation frequency is varied here to identify resonant responses. The resonance peaks shift with $D$, consistent with the interaction-dependent effective fields and precessional dynamics. Their amplitudes also vary because the measured response depends on the coupling of the applied field to each collective coordinate, as well as on damping and the definition of the plotted amplitude. The symmetric and antisymmetric spectra distinguish the combined and relative responses, whereas the first-spin spectrum alone does not uniquely identify the dominant collective channel. For comparison, the analytical linearized frequencies corresponding to symmetric and antisymmetric modes at $h_{\mathrm{dc}}+K=1.10$ are given by Eq.~(\ref{omega_s}) and (\ref{omega_a}):
\begin{align}
	\omega^{}_S &= \sqrt{1.10(1.10-3D)}\\
	\omega^{}_A &= \sqrt{(1.10-2D)(1.10+D)}
\end{align}
The above analytical formulae  demonstrate interaction-dependent resonance responses and their distribution between collective coordinates in the linear response regime. It is also clear from the Fig.~(\ref{Figure11}) that the simulated FMR response is consistent with analytically obtained angular frequency $\omega^{}_S$ and $\omega^{}_A$ in linear response regime namely $D\le0.367$ (For symmetric mode) and $D\le0.55$ (For antisymmetric mode), respectively.

\begin{figure*}
	\centering\includegraphics[width=0.75\textwidth]{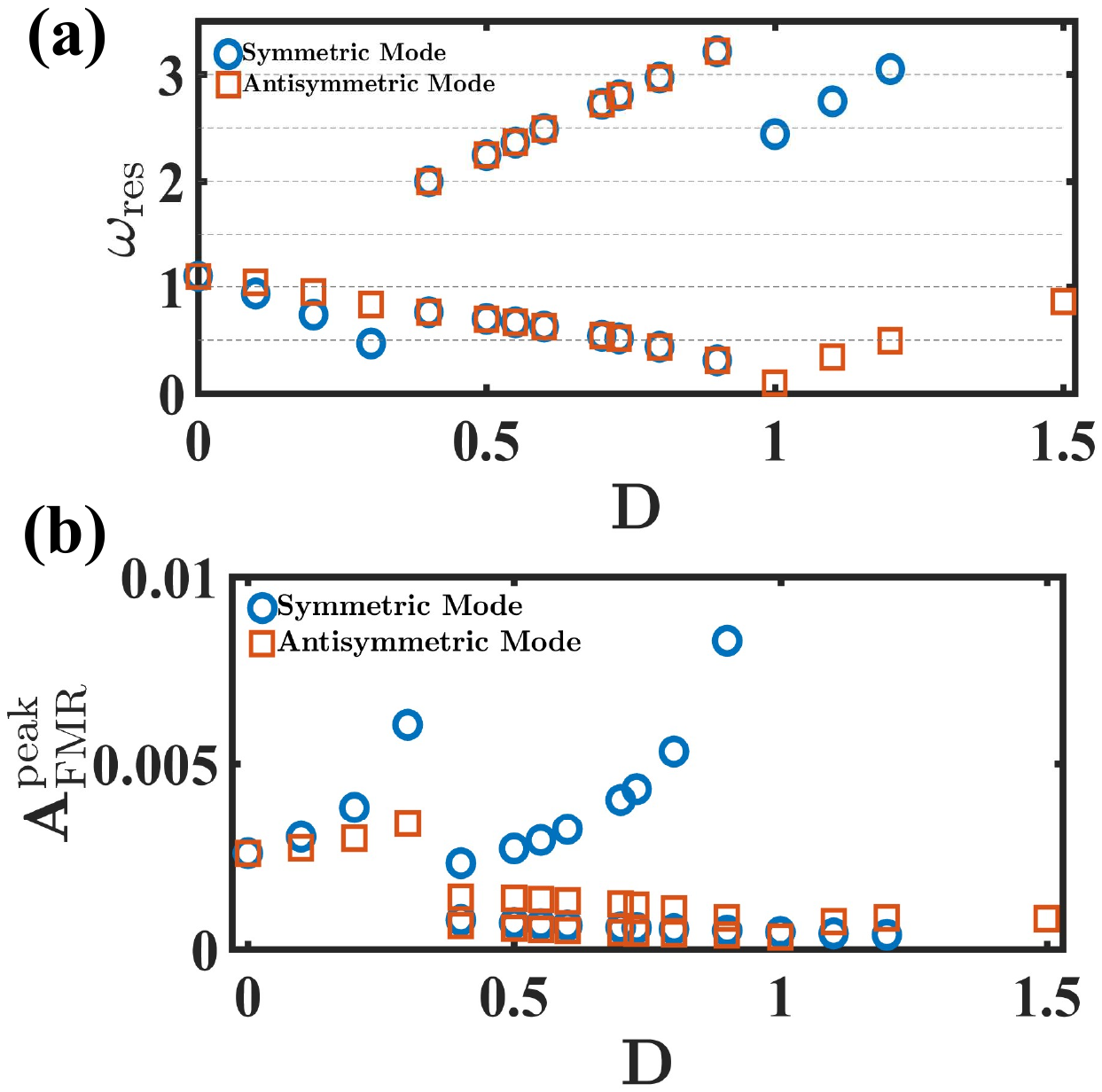}
	\vspace{-0.5cm}
	\caption{Dependence of the collective-mode FMR characteristics on dipolar coupling strength $D$. (a) Resonance angular frequencies of the symmetric and antisymmetric modes. (b) Corresponding peak FMR amplitudes.}
	\label{Figure12}
\end{figure*}
Fig.~(\ref{Figure12}) summarizes the FMR spectra by extracting the resonance frequencies and peak amplitudes of the symmetric and antisymmetric responses. In panel (a), the two resonance branches show different dependences on $D$, consistent with the distinct coupling dependence of the linearized collective modes. Dipolar coupling therefore changes both the characteristic precessional timescales and the relative positions of the collective resonances. In panel (b), the peak amplitudes vary non-monotonically with $D$. A shift in resonance frequency does not necessarily produce a larger peak amplitude: the peak height also depends on how strongly the applied field excites the mode, the damping, and the response amplitude used in the analysis. Consequently, the frequency shift and peak-height variation represent distinct aspects of the FMR response and should not be interpreted interchangeably. Taken together, the two panels summarize how dipolar coupling reorganizes both the characteristic frequencies and the relative strengths of the collective responses.}

%%%%%%%%%%%%%%%%%%%%%%%%%%%%%%%%%%%%%%%%%%%%%%%%%%%%%%%%%%%%%%%%%%%%%%

\section{Summary and Discussion}
In summary, we have investigated the non-linear dynamics of two dipolar-coupled anisotropic macrospins governed by the Landau–Lifshitz–Gilbert equation under periodic transverse excitation. Using bifurcation diagrams, Poincaré sections, Lyapunov exponents, Fourier spectra, and ferromagnetic-resonance analysis, we have examined how dipolar coupling, transverse driving amplitudes, and uniaxial anisotropy jointly govern the stability and long-time dynamics of the system. 
The results presented in the present work establish that the non-linear response of the driven two-spin system is governed by the interplay between dipolar coupling, uniaxial anisotropy, and periodic transverse excitation. This interplay is rooted in the structure of the effective field itself. For the geometry considered here, the dipolar contribution is not isotropic: its strength is $2D$ along the $x$ direction and $-D$ along the $y$ and $z$ directions. Consequently, changing $D$ does more than increase the strength of the interaction between the spins. It changes the directional balance of the effective fields, and therefore modifies the precessional torque experienced by each moment as its orientation evolves. Because the LLG equation couples all magnetization components through the cross product and damping terms, this directional modification feeds back into the complete trajectory. The resulting dynamics cannot, in general, be understood by treating the dipolar interaction as a simple shift of a single resonance frequency.

The collective coordinates provide a useful way to interpret this coupled motion. The symmetric coordinate, $\mathbf m_S=(\mathbf m_1+\mathbf m_2)/\sqrt{2}$, describes the combined or predominantly in-phase response, whereas the antisymmetric coordinate, $\mathbf m_A=(\mathbf m_1-\mathbf m_2)/\sqrt{2}$, tracks the relative difference between the two spins. These coordinates do not merely re-express the individual magnetizations; they help distinguish changes in the net collective response from changes in the degree to which the two spins move differently. The coupling-dependent structures observed in the symmetric and antisymmetric components indicate that the two-spin response involves both types of motion. 
%In particular, a change in the collective signal need not be accompanied by an equivalent change in the relative-spin signal. This distinction is physically important because the interaction acts on each spin through the orientation of the other, allowing small differences between the two trajectories to alter the subsequent torques and, in turn, the evolution of their difference.

The non-monotonic variation of the dynamical response with $D$ is consistent with this competition between directional dipolar torques, anisotropy, and the external drive. As the coupling is varied, the Poincar\'e-sampled magnetization branches change from relatively organized structures to split or densely populated regions, while intervals of more organized response can reappear. The local-maxima-based bifurcation diagram show a corresponding reorganization of the continuous-time oscillations. Such behaviour is not consistent with a simple picture in which increasing interaction strength steadily increases the complexity of the dynamics. Instead, the interaction changes the stability and accessible form of the driven trajectories in a parameter-dependent manner. Branch splitting and broad distributions are useful indicators of changes in the long-time response, but they do not alone identify the mechanism of a transition: periodic-orbit bifurcations, higher-period responses, quasi periodicity, and chaotic motion can produce some overlapping visual features. Their interpretation therefore relies on the combined evidence from the Poincar\'e sections, Local maxima, and Lyapunov analysis.

The Lyapunov exponents add a stability-based interpretation to these geometric observations. A positive largest Lyapunov exponent indicates exponential separation of nearby trajectories and, when sustained after the transient and supported by the other diagnostics, provides evidence for deterministic chaos. The alternating intervals of positive and near-zero leading exponents show that the driven system does not remain in one dynamical regime throughout the coupling sweep. Rather, changes in $D$ can enhance or suppress sensitivity to initial conditions. This is an important distinction from simply observing a large-amplitude or irregular magnetization signal: amplitude alone does not determine dynamical instability. Similarly, the simultaneous positivity of the two largest exponents is consistent with hyperchaos. The mode-oriented growth rates provide complementary information about the response to perturbations initially directed along symmetric or antisymmetric coordinates. 
%but they should not automatically be regarded as independent mode-specific Lyapunov spectra unless those perturbation sectors remain invariant during the evolution. This qualification is especially relevant when using collective coordinates to interpret instability in a system whose non-linear trajectory may not preserve exact spin-exchange symmetry.

The drive-amplitude dependence reinforces the conclusion that the response is controlled by the geometry of the effective field, not only by the overall strength of excitation. Although $h_x$ and $h_y$ have the same sinusoidal time dependence, they act along different spatial directions. In the present model, the applied field is linearly polarized along a fixed transverse direction determined by the ratio of its two components; it is not a circularly rotating field. Moreover, the dipolar field distinguishes the $x$ and $y$ directions through its unequal coefficients. Varying $h_x$ therefore changes the torque balance differently from varying $h_y$, even when the temporal waveform is unchanged. The different parameter-dependent responses found in the two sweeps are consequently physically reasonable: each drive component changes the orbit followed by the spins and the way that orbit samples the anisotropic interaction field. The results also show why stronger excitation should not be assumed to produce a uniform increase in chaos. A larger drive can shift the parameter intervals in which complicated dynamics occur, while leaving other intervals comparatively regular. 
%The relevant quantity is the combined effect of drive, coupling, and anisotropy on the trajectory and its stability.

Uniaxial anisotropy contributes another competing field scale. Since its effective-field contribution is proportional to $K m_{iz}$ along the easy axis, its influence depends on the instantaneous out-of-axis orientation of each spin. Varying $K$ changes the restoring torque and hence the precessional response, but its effect is mediated by the dipolar field and the periodic forcing. This explains why the dependence on $K$ is non-monotonic and why the response for uncoupled and coupled spins can differ substantially. In the uncoupled limit, each spin responds to the anisotropy and external drive without a dipolar field from the other spin; at finite coupling, the orientation of one spin contributes directly to the effective field of the other. The anisotropy sweep therefore probes a balance among the easy-axis restoring tendency, interaction-induced torques, and the time-dependent drive. 
%It is not appropriate to classify anisotropy as universally stabilizing or destabilizing from these results; its influence depends on the values of the other parameters.

The two-dimensional Lyapunov maps extend this interpretation by showing that stability boundaries depend on combinations of control parameters. In the $D$--$h_x$ and $D$--$h_y$ planes, the curved bands and structured regions imply that a single critical drive amplitude cannot describe the onset of instability for all coupling strengths. Likewise, the $D$--$K$ and $h_x$--$h_y$ maps demonstrate that the stability landscape depends on the joint parameter values rather than on one parameter in isolation. These maps are particularly useful for identifying regions where perturbation growth is enhanced or suppressed, but their contours should not be assigned to a specific bifurcation mechanism without additional continuation or stability analysis. The maps also make clear that the two transverse drive components are not interchangeable: their influence is filtered through the anisotropic dipolar geometry, even though their time dependence is identical.

The Fourier spectra and magnetization trajectories offer complementary views of the same non-linear response. The spectra show how the response is distributed across frequencies and how this distribution changes with coupling. The presence of harmonics is compatible with non-linear LLG dynamics even though the applied field contains a single sinusoidal frequency; the non-linear torque terms can generate components that are absent from the forcing itself. Changes in relative peak heights and in the weight carried by additional frequencies indicate that $D$ modifies the temporal structure of the magnetization motion. Comparing the first-spin spectrum with the symmetric and antisymmetric spectra helps distinguish frequencies associated predominantly with the combined response from those associated with differences between the spins. 
%A very small antisymmetric spectral amplitude should be interpreted as a weak relative-spin response on the plotted scale, not as a divergence or, by itself, as evidence of stability. Conversely, a broad spectrum alone does not prove chaos, because spectral broadening must be interpreted together with the dynamical and Lyapunov diagnostics.

The three-dimensional trajectories provide a geometric representation of how the spins and collective coordinates explore their state spaces. Restricted, repeated curves are compatible with periodic motion, whereas denser or more space-filling-looking projections indicate a more complicated sampled trajectory. The individual-spin plots show how each moment evolves, while the collective-coordinate plots reveal whether the complexity is expressed in the combined response, the relative response, or both. Nevertheless, these plots are projections of the dynamics and can obscure structure; a visually dense curve is not a sufficient criterion for chaos. Their value lies in their consistency with the bifurcation, spectral, and Lyapunov results, rather than in serving as an independent proof of a chaotic attractor.

The analytical linearization provides a separate reference for understanding the frequency-dependent response. Around the collinear state $\mathbf m_1=\mathbf m_2=\hat{\mathbf z}$ with the AC field absent, the collective coordinates obey linear oscillator equations with symmetric and antisymmetric angular frequency $\omega^{}_S$ and $\omega^{}_A$ given by Eq.~\ref{omega_s} and Eq.~\ref{omega_a}, respectively. For the parameter choice $h_{dc}+K=1.10$, these reduce to $\omega^{}_S=\sqrt{1.10(1.10-3D)}$ and $\omega^{}_A=\sqrt{(1.10-2D)(1.10+D)}$. The expressions show that dipolar coupling affects the two collective modes differently and can soften their characteristic frequencies at different coupling values. The symmetric-mode angular frequency reaches zero at $D\approx0.3667$, while the antisymmetric-mode angular frequency reaches zero at $D\approx0.55$. Beyond the respective thresholds, the corresponding linearized expression no longer gives a real oscillatory frequency about the assumed collinear reference state. This is a limitation of that reference-state linearization.
% not a claim that a numerical resonance must vanish or that the full non-linear system ceases to evolve. In particular, the analytical expressions should not be extrapolated as real resonance frequencies into regions where the relevant squared frequency is negative.

The numerical FMR spectra address a different question from the fixed-frequency non-linear scans. By varying the excitation frequency, we have identified resonant features of the driven response, while the non-linear scans examine how the long-time dynamics changes as a parameter is varied at fixed drive frequency. The observed coupling-dependent resonance positions are consistent with the fact that the dipolar interaction modifies the effective fields and the characteristic precessional timescales. The symmetric and antisymmetric responses need not have identical resonance positions or amplitudes because the two collective coordinates respond differently to the interaction and to the applied field. Peak amplitude is also distinct from resonance frequency: it depends on the excitation of the relevant response, damping, and the amplitude measure used in the analysis. The linearized frequencies provide a useful quantitative reference where the collinear state is an appropriate approximation, whereas the numerical spectra reflect the response under the specific excitation and measurement protocol.

In conclusion, the results shown in the present work show that this minimal two-spin system supports a rich, parameter-dependent range of non-linear responses. The physical origin of this richness is the coupling of the directionally anisotropic dipolar field, combined with uniaxial anisotropy, damping, and periodic transverse excitation. The collective coordinates reveal how the response separates into combined and relative motion; the Poincar\'e sections and local maxima characterize the organization of the long-time dynamics; the Lyapunov exponents quantify sensitivity to initial conditions; and the Fourier spectra, trajectory, and FMR analyses provide frequency-domain, geometric, and linear-response perspectives. The agreement among these complementary diagnostics supports the interpretation of deterministic chaotic regimes while also identifying the limits of what any single diagnostic can establish. The central implication is that non-linear instability is not set by dipolar coupling, drive amplitude, or anisotropy independently, but emerges from their combined action on the effective fields and the stability of the driven trajectories.
\newpage
\bibliography{ref}
\end{document}